\documentclass[prd,a4paper,nofootinbib,final,twocolumn,floatfix]{revtex4}
\usepackage{amsfonts}
\usepackage{amsmath}
\usepackage{amssymb}
\usepackage[usenames,dvips]{color}
\usepackage[english]{babel}
\usepackage[latin1]{inputenc}
\usepackage{amsfonts}
\usepackage{appendix}
\usepackage{epsfig}
\usepackage{graphics,rotating}
\usepackage{dcolumn}
\usepackage{bm}
\usepackage{color}
\usepackage[colorlinks=true,
            linkcolor=blue,
            urlcolor=black,
            citecolor=blue]{hyperref}
\usepackage{hyperref}
\usepackage[T1]{fontenc}
\usepackage{float}
\usepackage{subfigure}

\begin{document}

\title{Geometrical and physical interpretation of the Levi-Civita spacetime 
\\
in terms of the Komar mass density}
\author{Bence Racskó$^{\dag }$, László Árpád Gergely$^{\star }$}
\affiliation{Institute of Physics, University of Szeged, D\'om t\'er 9, Szeged 6720,
Hungary\\
$^{\dag }${\small E-mail: racsko@titan.physx.u-szeged.hu\quad \quad }$%
^{\star }${\small \ E-mail: laszlo.a.gergely@gmail.com }}

\begin{abstract}
We revisit the interpretation of the cylindrically symmetric, static vacuum
Levi-Civita metric, known in either Weyl, Einstein-Rosen, or Kasner-like
coordinates. Despite the infinite axis source, we achieve a rigurous
derivation of its Komar mass density through a compactification and
subsequent blowing up of the compactification radius. The Komar mass density
turns out coordinate system dependent, the timelike Killing vector (in the
absence of a preferred normalization) being adopted to the coordinate time
in each of the available systems. We show that among all possibilities, the
Komar mass density $\mu _{K}$ calculated in the Einstein-Rosen frame, when
employed as the metric parameter, has a number of advantages over other
parametrizations. First it eliminates previous double coverages of the
parameter space. It vanishes in flat spacetime and when small, it
corresponds to the mass density of an infinite string. It characterizes the
Kretschmann scalar in the simplest possible way (no double coverage of the
parameter space, neither a multivalued function, as with other parameters).
After a comprehensive analysis of the local and global geometry (including a
study of the singularity on the axis, based on the Królak and Tipler
criteria, and refuting earlier related claims), we proceed with the physical
interpretation of the Levi-Civita spacetime. First we show that the
Newtonian gravitational force is attractive and its magnitude increases
monotonically with all positive $\mu _{K}$, asymptoting to $R^{-1}$ (with $R$
the proper distance in the "radial" direction). Second, we reveal that the
magnitude of the tidal force between nearby timelike geodesics (hence
gravity in the Einsteinian sense) attains a maximum at $\mu _{K}=1/2$ and
then decreases asymptotically to zero. Hence, from a physical point of view
the Komar mass density of the Levi-Civita spacetime encompasses two
contributions: Newtonian gravity and acceleration effects. An increase in $%
\mu _{K}$ strengthens Newtonian gravity but also drags the field lines
increasingly parallel, eventually transforming Newtonian gravity through the
equivalence principle into a pure acceleration field (with acceleration $%
R^{-1}$) and the Levi-Civita spacetime into a flat Rindler-like spacetime.
In a geometric picture the increase of $\mu _{K}$ from zero to $\infty $
deforms the $t=$const, $z=$const planar sections of the Levi-Civita
spacetime into ever deepening funnels, eventually degenerating into
cylindrical topology.
\end{abstract}

\maketitle

\section{Introduction}

Gravitational waves were predicted in the early days of general relativity,
as wavelike perturbations of flat spacetime propagating with the speed of
light \cite{GWfirst}, but only in recent years they were detected (see the
catalog of 90 gravitational waves \cite{GWTC3} by the LIGO Scientific, Virgo
and KAGRA Collaborations). As the existence of spherically symmetric
gravitational waves in vacuum is forbidden by the Jebsen-Birkhoff theorem,
the next simplest geometry would be cylindrically symmetric. Cylindrical
symmetry also provides the next simplest setup (after spherical symmetry)
for discussing both mathematical and physical aspects of spacetimes with
localized sources, emerging as solutions of the Einstein equation and it is
also a precursor for studying axial symmetry.

Unlike in the spherically symmetric case, the cylindrically symmetric vacuum
is not unique. The Einstein-Rosen cylindrically symmetric vacuum solutions
include wavelike behaviors, allowing for both standing wave and approximate
progressive wave solutions, discovered analytically in the very early days
of general relativity by Einstein and Rosen \cite{EinsteinRosen}. Later both
solitonic waves \cite{BelinskiiZakharov} and impulsive wave solutions \cite%
{Carmeli}, \cite{HerreraSantos} were identified in this class.

The canonical quantization of cylindrically symmetric gravitational waves by
Kucha\v{r} was the earliest example of the midisuperspace approach \cite%
{KucharCylindric}, which encompasses a much richer structure than previous
minisuperspace quantizations of the Friedmann and mixmaster universes (by
DeWitt \cite{DeWitt} and Misner \cite{Misner1,Misner2,Misner3},
respectively). Due to a fair compromise between the simplicity induced by
degrees of freedom frozen by symmetry assumptions on the one hand and the
full complexity of the gravitational degrees of freedom on the other hand,
cylindrical gravitational waves provide an ideal testbed for comparing
quantization approaches.

In a strong field a fast changing gravitational wave can be separated from a
slowly changing background in the geometrical optics (high frequency)
approximation, as discussed thoroughly by Isaacson \cite{Isaacson1,Isaacson2}%
. The key concept here is that the curvature radius of the background ought
to be much larger than the wavelength. A cylindrically symmetric, static
background for the cylindrical gravitational waves, which could represent
such a strong field, is the Levi-Civita spacetime \cite{Levi-Civita}, the
static limit of the Einstein-Rosen class.

This static vacuum solution was derived by Levi-Civita in 1919 \cite%
{Levi-Civita}, with the intent to characterize the gravitational vacuum
outside a cylindrically symmetric source\textbf{.} Generalizations of the
static Levi-Civita solution beyond vacuum are also known. The exterior of a
radiating cylinder was discussed in Ref. \cite{Rao}. This radiating
Levi-Civita spacetime contains a null dust, in this sense being similar to
the spherically symmetric Vaidya metric. Electrovacuum generalisations also
emerged, including the inclusion of axial or longitudinal magnetic fields by
Bonnor \cite{Bonnor} and radial electric field by Raychaudhuri \cite%
{Raychaudhuri}.

The Levi-Civita static and Einstein-Rosen wavelike solutions with their
possible sources have been studied in Refs. \cite{Marder}, \cite{Thorne}, 
\cite{Bonnor79}, \cite{GriffithsP}. However, even for the static vacuum case
certain aspects related to its physical interpretation remain elusive. It is
the purpose of this paper to revisit the static Levi-Civita spacetime,
sheding light on these key properties, both from geometric and physical
viewpoints.

The Levi-Civita metric was discovered and has been often discussed in the
Weyl form, suitable for static and axially symmetric scenarios. While a
small positive metric parameter $\lambda $~has the convenient interpretation
of the linear mass density along the symmetry axis of the cylinder (it
agrees with it in a first order expansion in $\lambda $ \cite{Marder}), the
interpretation seems to break down at $\lambda =1/4$, above which the
Kretschmann scalar decreases with increasing $\lambda $ \cite%
{GautreauHoffman}, \cite{Bonnorlambda4}. Furthermore, there is no unique
flat spacetime limit, the Levi-Civita spacetime becoming flat for the
quartet of values $\lambda =0$, $1/2$ and\textbf{\ }$\lambda \rightarrow \pm
\infty $ \cite{GriffithsP}. Clearly, the metric parameter $\lambda $ lacks a
crystal clear physical meaning, being by far no proper analogue for the mass
parameter of the Schwarzschild spacetime.

Einstein-Rosen waves are naturally described in another coordinate system,
suitable for cylindrical symmetry. This is the particular nonrotating limit
of the Jordan-Ehlers-Kundt-Kompaneets coordinates \cite{Bini}. The Weyl and
Einstein-Rosen metric forms are related through the analytical continuation $%
t\rightarrow i\widehat{z}$, $z\rightarrow i\widehat{t}$ (with $t$ and $z$
the temporal and axial Einstein-Rosen coordinates, their Weyl counterparts
carrying a hat), generating an axially symmetric stationary metric in the
Weyl form from a cylindrically symmetric metric. The Levi-Civita spacetime,
being both static and cylindrically symmetric, in addition allows for a real
coordinate transformation between these two forms \cite{Marder}, \cite%
{Thorne}.

In Sec. 2 we summarize the preliminaries necessary for our discussion. We
start from the Einstein-Rosen form (also known as canonical, derived in
Appendix \ref{stacan} from a more general standard form) of a cylindrically
symmetric metric with vorticity-free Killing vectors and orthogonally
transitive group action, to specify the metric functions leading to the
Levi-Civita metric. We present the transformation between its Weyl and
Einstein-Rosen forms explicitly, as we were unable to locate it elsewhere in
the literature. The Levi-Civita metric in the Einstein-Rosen form closely
mimics the properties established in Weyl coordinates. For a quartet of
values of the metric parameter $\sigma =0$, $1$, and $\pm \infty $ the
curvature disappears. For small negative values of the parameter $\sigma $
the interpretation of a (Newtonian) gravitational field generated by a
homogeneous cylinder holds. For other values however, it breaks down,
including a range of parameter values with repulsive gravity. For both the
Weyl and the Einstein-Rosen forms the properties of the respective
parameters suggest a double coverage of the available configurations. At the
end of Sec. 2., preparing for the next section, we summarize the essentials
of the Komar superpotential and charges.

Sec. 3. contains a rigorous derivation of the Komar mass density $\mu _{K}$
of the Levi-Civita spacetime in the Einstein-Rosen coordinates. This is
achieved despite the infinite source along the axis, through a
compactification and subsequent blowing up of the compactification radius.
Next we propose $\mu _{K}$ as yet another parameter of the Levi-Civita
metric, which has the advantage to eliminate the double coverage encountered
before. The metric is flat for $\mu _{K}=0$ and the Komar mass density has
the interpretation of linear mass density on the symmetry axis up to $\mu
_{K}=1$. At $\mu _{K}\rightarrow \infty $ the metric is flat again, this
time in uniformly accelerated Rindler coordinates. As with increasing $\mu
_{K}$ the Levi-Civita metric approaches the Rindler limit, we conjecture
that beside mass and Newtonian gravitational energy, the Komar mass density
also encompasses acceleration contributions. We illustrate this point in
Appendix \ref{KomarR} showing that the Rindler metric also generates a
nonvanishing Komar mass density. At the end of the section we discuss the
C-energy \cite{Thorne}, \cite{Bini} of the Levi-Civita metric, showing that
it increases monotonically with $\mu _{K}\geq 0$. When reexpressing it in
terms of the Komar mass density calculated in the Weyl coordinates, $\left(
\mu _{K}\right) _{\mathrm{Weyl}}=$ $\lambda $, this property is lost,
supporting the claim that $\mu _{K}$ is the best available metric parameter.

In Sec. 4. we analyze the behaviour of the curvature invariants parametrized
by the Komar mass density, in terms of the proper radial distance. We find a
maximum of the Kretschmann scalar $\mathcal{K}$ at $\mu _{K}=1$ and the rest
of the scalars expressed in terms of $\mathcal{K}$ and $\mu _{K}$ alone. The
increase and subsequent decrease of the Kretschmann curvature with $\mu _{K}$
is counterintuitive, undermining the interpretation of $\mu _{K}$ as mass
density for $\mu _{K}>1$. Despite the metric apparently diverging for $\mu
_{K}\rightarrow \infty $, the curvature invariants vanish there (which comes
as no surprise as $\mu _{K}\rightarrow \infty $ corresponds to $\lambda =1/2$%
, where the Levi-Civita spacetime is known to be flat). This emerging
flatness is manifest in the well-known Kasner-like coordinate system \cite%
{GriffithsP}, which is regular for both flat limits and explores a redefined
time together with the proper radial distance as new coordinates. We rewrite
the Kasner parameters in terms of the Komar mass density. For $\mu
_{K}\rightarrow \infty $ the metric emerges flat in accelerated Rindler
coordinates.

Recently Ref. \cite{Ahmed} presented a new cylindrically symmetric vacuum
spacetime, claiming that the axis is a null geodesically incomplete soft
singularity both in the sense of Królak \cite{Krolak} and of Tipler \cite%
{Tipler}. The C-energy density measured by an observer was also computed,
the algebraic type shown to be Petrov type D and the geodesic deviation of
timelike geodesics synchronized by the proper time. In Appendix \ref{AhmedNO}
we prove that the metric of Ref. \cite{Ahmed}\ is nothing but the particular
case of the Levi-Civita metric for $\mu _{K}=1$. By taking the particular
case of the curvature invariants we correct the respective expressions of
Ref. \cite{Ahmed}. Then in the last subsection of Sec. 4. we analyse the
nature of the singularity on the symmetry axis and prove that the radial
null geodesics in the Levi-Civita spacetime obey both the Królak and the
Tipler strong singularity conditions, irrespective of the value of $\mu _{K}$
(hence we disprove the claim of Ref. \cite{Ahmed}, according to which the
singularity on the axis ought to be soft). The technicalities of the proof
are deferred to Appendix \ref{StrongSing}.

In Sec. 5. we proceed with the physical interpretation of the Levi-Civita
spacetime for generic Komar mass densities by considering the acceleration
necessary to keep a stationary observer in orbit at fixed proper distance
from the axis. We show that for positive $\mu _{K}$ the gravitational
acceleration is attractive and increases monotonically with $\mu _{K}$,
asymptoting to a constant value. Hence, despite increasing $\mu _{K}$,
Newtonian gravitational attraction cannot increase above a certain limit.
Then, we study the magnitude of the tidal forces, showing that the geodesic
deviation also exhibits a maximum, this time at $\mu _{K}=1/2$.

We summarize the geometric and physical characterization of the Levi-Civita
metric in the discussion presented as Sec. 6.

Throughout the paper we use units $G=1=c$.

\section{Preliminaries}

In this section we summarize the main ingredients necessary for a subsequent
thorough investigation of the Levi-Civita spacetime.

\subsection{Einstein-Rosen and Weyl forms of the Levi-Civita spacetime}

The Einstein--Rosen, or canonical form of the line element of a generic
vacuum cylindrically symmetric spacetime with vorticity-free Killing vectors
and orthogonally transitive group action (dubbed as whole-cylinder symmetry
by Thorne \cite{Thorne}), is (for a derivation see Appendix \ref{stacan}) 
\begin{equation}
\mathrm{d}s^{2}=e^{2\left( K-U\right) }\left( -\mathrm{d}t^{2}+\mathrm{d}%
r^{2}\right) +e^{-2U}r^{2}\mathrm{d}\varphi ^{2}+e^{2U}\mathrm{d}z^{2}~,
\label{canonical}
\end{equation}%
with $K$ and $U$ functions of the coordinates $\left( t,r\right) $. Here all
coordinates are dimensionless. For $K=0=U$ or $K=U=\ln r$ the line element (%
\ref{canonical}) degenerates into the flat metric.

These symmetries and the vacuum condition do not guarantee a unique solution
of the Einstein equations. Indeed, they allow for various type of
Einstein--Rosen waves \cite{EinsteinRosen} beside the static Levi-Civita
solution. The latter emerges for\footnote{%
See Eq. (3) of Ref. \cite{Rao} or Eqs. (4), (28), (29) of Ref. \cite%
{AkyarDelice}, with $A=0$, $W=r$ and $k=1$.} 
\begin{equation}
U=\sigma \ln r~,\quad K=\sigma ^{2}\ln r~,  \label{UK}
\end{equation}%
giving%
\begin{equation}
\mathrm{d}s^{2}=r^{2\sigma \left( \sigma -1\right) }\left( -\mathrm{d}t^{2}+%
\mathrm{d}r^{2}\right) +r^{2\left( 1-\sigma \right) }\mathrm{d}\varphi
^{2}+r^{2\sigma }\mathrm{d}z^{2}~,  \label{LC}
\end{equation}%
with a constant $\sigma \in \mathbb{R}$.

Levi-Civita considered a different line element in the axially symmetric and
static Weyl form:%
\begin{equation}
\mathrm{d}s^{2}=-\widehat{r}^{4\lambda }\mathrm{d}\widehat{t}^{2}+\widehat{r}%
^{4\lambda \left( 2\lambda -1\right) }\left( \mathrm{d}\widehat{r}^{2}+%
\mathrm{d}\widehat{z}^{2}\right) +\widehat{r}^{2\left( 1-2\lambda \right) }%
\mathrm{d}\widehat{\varphi }^{2}~,  \label{eq:LC_W}
\end{equation}%
with $\lambda \in \mathbb{R}$. Both metrics (\ref{LC}) and (\ref{eq:LC_W})
are static and cylindrically symmetric, hence they ought to be related.
Indeed, they transform into each other through the analytical continuation $%
t\rightarrow i\widehat{z}$, $z\rightarrow i\widehat{t}$ \cite{ExSol}.

An explicit coordinate transformation%
\begin{align}
t& =\left( \sigma -1\right) ^{-2p_{0}}\widehat{t}~,  \notag \\
r& =\left( \sigma -1\right) ^{2/\left( \sigma ^{2}-\sigma +1\right) }%
\widehat{r}^{1/\left( \sigma -1\right) ^{2}}~,  \notag \\
\varphi & =\left( \sigma -1\right) ^{-2p_{+}}\widehat{\varphi }~,  \notag \\
z& =\left( \sigma -1\right) ^{-2p_{-}}\widehat{z}~,  \label{ERWtraf}
\end{align}%
with%
\begin{equation}
p_{0}=\frac{\sigma \left( \sigma -1\right) }{\sigma ^{2}\!-\!\sigma \!+\!1}%
~,~~p_{+}=\frac{1-\sigma }{\sigma ^{2}\!-\!\sigma \!+\!1}~,~~p_{-}=\frac{%
\sigma }{\sigma ^{2}\!-\!\sigma \!+\!1}  \label{p0pm}
\end{equation}%
can also be constructed, where%
\begin{equation}
\sigma =\frac{2\lambda }{2\lambda -1}~.  \label{paramtransf}
\end{equation}%
The coordinate transformation (singular for $\sigma =1$ or $\lambda =1/2$)
is obtained by transforming both metrics into a Kasner-like form. In the
case of a small parameter $\lambda \approx -\sigma /2$ (thus $\sigma $ also
small), (\ref{ERWtraf}) is close to the identity transformation and reduces
to it for $\sigma \rightarrow 0$.

The line elements (\ref{LC}) and (\ref{eq:LC_W}) are thus locally isometric%
\textbf{.} The isometry fails to be global if both $\varphi $ and $\widehat{%
\varphi }$ are angular coordinates with period $2\pi $. Indeed, if the
Einstein-Rosen azimuthal angle is periodic with $2\pi $, the period of the
Weyl azimuthal coordinate becomes $2\pi \left( \sigma -1\right) ^{2p_{+}}$.

For $0\leq \lambda \ll 1$ (thus $\sigma \approx -2\lambda $ in the small
negative value range) the parameter $\lambda $ has the interpretation of the
constant mass per unit length of a static cylinder with negligible internal
pressure \cite{Thorne}. Indeed, the cylindrically symmetric Laplace equation%
\begin{equation}
\nabla ^{2}\phi _{N}=\frac{1}{r}\frac{d}{dr}\left( r\frac{d\phi _{N}}{dr}%
\right) =0
\end{equation}%
for the Newtonian potential is solved as $\phi _{N}=2m\ln r$ (with $m$ an
integration constant and another irrelevant integration constant dropped).
Defining the mass $M=\int_{V}\rho dV$ of a cylindrical volume $V$
(containing a distributional source $\rho $ on the axis) through the Poisson
equation $\nabla ^{2}\phi _{N}=4\pi \rho $ leads to 
\begin{equation}
M=\frac{1}{4\pi }\int_{\partial V}\mathbf{\nabla }\phi _{N}\cdot d\mathbf{A}%
=m\int dz\mathbf{~,}
\end{equation}%
where $d\mathbf{A}$ is the outward directed normal of the boundary $\partial
V$ of the cylinder (however as $\mathbf{\nabla }\phi _{N}=(2m/r)\mathbf{e}%
_{r}$, only the cylindrical surface contributes). Hence, the constant $m$ is
precisely the mass density along the $z$-axis. In the\textbf{\ }stationary,
weak field and slow motion limit $g_{00}\approx -1-2\phi _{N}=-1-4m\ln r$.
On the other hand, the Levi-Civita metric in Einstein-Rosen coordinates has $%
g_{00}=-r^{2\sigma \left( \sigma -1\right) }=-\exp \left[ 2\sigma \left(
\sigma -1\right) \ln r\right] $, which for small $\sigma $ approximates as $%
g_{00}\approx -1-2\sigma \left( \sigma -1\right) \ln r\approx -1+2\sigma \ln
r\approx -1-4\lambda \ln r$, confirming the interpretation $\lambda \approx
m $, when small.

In deriving the solution (\ref{UK}) a constant of integration was supressed
through the requirement to recover the Minkowski metric in cylindrical
coordinates when $\lambda =0$ (hence $\sigma =0$). For $\sigma =1$ \textbf{[}%
hence\textbf{\ }$\lambda \rightarrow \pm \infty $, due to Eq. (\ref%
{paramtransf})\textbf{]} the metric is also flat (although some of the
metric coefficients diverge, other vanish). Clearly, the interpretation of $%
\lambda $ as mass per length is unsuitable for its whole range. In fact
there is a duality in the ranges of either parameters $\lambda $ or $\sigma $%
, corresponding to an interchange of the coordinates $z$ and $\varphi $ \cite%
{GriffithsP}. This duality appears as an unnecessary redundancy in either of
the parametrizations.

Moreover, the metric (\ref{LC}) diverges for $\sigma \rightarrow \pm \infty $%
. In this case, however it can be transformed to the flat metric perceived
by an accelerated observer, the Rindler metric. This is exactly the flatness
of the Levi-Civita metric in the Weyl form emerging for $\lambda =1/2$ \cite%
{GriffithsP}, see Eq. (\ref{paramtransf}).

\subsection{The Komar superpotential and charges}

We consider a vector field $\xi $ and corresponding $1$-form $\boldsymbol{%
\xi }$ on a four dimensional Lorentzian spacetime $M$. Its Komar
superpotential\emph{\ }$2$-form \cite{Komar}%
\begin{equation}
\mathbf{U}_{\xi }=\ast \mathrm{d}\boldsymbol{\xi }=\frac{1}{2}\left( \nabla
^{i}\xi ^{j}-\nabla ^{j}\xi ^{i}\right) \sqrt{\mathfrak{g}}\left( \mathrm{d}%
^{2}x\right) _{ij}~
\end{equation}%
(where $\mathfrak{g}=\left\vert \det g_{ij}\right\vert $ and $\left( \mathrm{%
d}^{2}x\right) _{ij}=\frac{1}{2}\epsilon _{ijkl}\mathrm{d}x^{k}\wedge 
\mathrm{d}x^{l}$) is defined through the Hodge dual 
\begin{equation}
\ast A=\frac{1}{4}A^{ij}\epsilon _{ijkl}\sqrt{\mathfrak{g}}\mathrm{d}%
x^{k}\wedge \mathrm{d}x^{l}
\end{equation}%
of the $2$-form $A=\mathrm{d}\boldsymbol{\xi }$, with $\mathrm{d}$ the
exterior derivative.

Then the current 
\begin{equation}
\mathbf{S}_{\xi }=\mathrm{d}\mathbf{U}_{\xi }=\nabla _{j}\left( \nabla
^{i}\xi ^{j}-\nabla ^{j}\xi ^{i}\right) \sqrt{\mathfrak{g}}\left( \mathrm{d}%
^{3}x\right) _{i}~
\end{equation}%
(with $\left( \mathrm{d}^{3}x\right) _{i}=\frac{1}{6}\epsilon _{ijkl}\mathrm{%
d}x^{j}\wedge \mathrm{d}x^{k}\wedge \mathrm{d}x^{l}$) is identically
conserved. This also emerges as the Noether current of the Einstein-Hilbert
action, associated to diffeomorphism invariance \cite{FF}.

When $\xi $ is a Killing vector field (hence $0=\nabla _{i}\xi _{j}+\nabla
_{j}\xi _{i}$), from the cyclic identity of the curvature tensor the
relation 
\begin{equation}
R_{\ ijk}^{l}\xi _{l}=\nabla _{i}\nabla _{j}\xi _{k}
\end{equation}%
emerges, which renders $\mathbf{S}_{\xi }$ into 
\begin{align}
\mathbf{S}_{\xi }& =2R_{\ j}^{i}\xi ^{j}\sqrt{\mathfrak{g}}\left( \mathrm{d}%
^{3}x\right) _{i}  \notag \\
& =16\pi \left( T_{\ j}^{i}-\frac{1}{2}T\delta _{j}^{i}\right) \xi ^{j}\sqrt{%
\mathfrak{g}}\left( \mathrm{d}^{3}x\right) _{i}~.
\end{align}%
In the last step we employed the Einstein equations. Hence in vacuum the
Komar superpotential of a Killing field is a closed 2-form.

If a closed set $\mathcal{C}\subseteq M$ encompasses all sources (including
four-dimensional extended sources, two-dimensional strings, one dimensional
point sourses, singularities, topological defects), then for a closed $2$%
-surface $\mathcal{S}\subseteq M\setminus \mathcal{C}$ the Komar charge 
\begin{equation}
Q_{\xi }\left( \mathcal{S}\right) =\int_{\mathcal{S}}\mathbf{U}_{\xi }
\end{equation}%
depends on the homology class of $\mathcal{S}$ only. Then every pair of
closed $2$-surfaces $\mathcal{S}$ and $\mathcal{S}^{\prime }$ encompassing $%
\mathcal{C}$ are homologous to each other, e.g. there is a $3$-surface $%
N\subseteq M\setminus \mathcal{C}$ such that $\partial N=\mathcal{S}-%
\mathcal{S}^{\prime }$. Stokes' theorem then gives 
\begin{equation}
Q_{\xi }\left( \mathcal{S}\right) -Q_{\xi }\left( \mathcal{S}^{\prime
}\right) =\int_{\partial N}\mathbf{U}_{\xi }=\int_{N}\mathrm{d}\mathbf{U}%
_{\xi }=0~,
\end{equation}%
which shows that the Komar charge is conserved. {If $\mathcal{S}$ and $%
\mathcal{S}^{\prime }$ are both spacelike and $N$ timelike, this corresponds
to a conservation law in the sense that }$Q_{\xi }${\ takes the same value
at all times.}

When $\xi $ is a timelike Killing vector field, then 
\begin{equation}
m_{K}=-\frac{1}{8\pi }Q_{\xi }\left( \mathcal{S}\right) =-\frac{1}{8\pi }%
\int_{\mathcal{S}}\mathbf{U}_{\xi }
\end{equation}%
is the Komar mass of the spacetime. The integral $m_{K}$ is however ambigous
to a constant factor, since $a\xi $ is an equally valid Killing vector (here 
$a\in \mathbb{R}^{+}$). The ambiguity can be removed when a preferred
normalization of the timelike Killing vector is available (like $\xi \cdot
\xi \rightarrow -1$ at asymptotic infinity).

\section{A new parametrization of the Levi-Civita metric}

In this section we first introduce a mathematically sound construction for
defining the Komar mass density $\mu _{K}$ for the Levi-Civita spacetime in
the Einstein-Rosen form, and will explore it as an alternative metric
parameter. One of its main advantages over the previously used parameters $%
\lambda $ or $\sigma $ is that it eliminates the double coverage appearing
in either of them. The $C$-energy of the spacetime provides additional
support for considering $\mu _{K}$ as a natural parameter of the Levi-Civita
spacetime.

\subsection{Komar mass density for the Levi-Civita spacetime}

The static, cylindrically symmetric Levi-Civita spacetime has a three
dimensional Killing algebra with a timelike Killing vector, a spacelike
axial Killing vector and a translational Killing vector along the $z$-axis
(which, being singular, is removed from the manifold). In the coordinates (%
\ref{LC}) the squared length of the timelike Killing vector field $\xi
=a\partial _{t}$ (with $a$ a constant) is $\xi \cdot \xi =-a^{2}r^{2\sigma
\left( \sigma -1\right) }$. Aside from special values of the parameter $%
\sigma =0,1$ where the metric is flat, this Killing vector field cannot be
normalized either at infinity or the axis, hence we adopt the simplest
choice $a=1$ (adapting the temporal Killing vector field to the coordinate
time).

The singularity on the axis $r=0$ extends to infinity, thus it is impossible
to wrap it in a closed $2$-surface, as required for the evaluation of the
Komar charge. Despite this in what follows we describe a procedure allowing
for the definition of the density $\mu _{K}$ of the Komar mass along the $z$%
-axis.

The key step is to compactify the $z$ direction as $z=l\alpha $, where $l$
is a length scale and $\alpha $ an angle parameter with periodicity $2\pi $.
The line element (\ref{LC}) becomes 
\begin{equation}
\mathrm{d}s^{2}=r^{2\sigma \left( \sigma -1\right) }\left( -\mathrm{d}t^{2}+%
\mathrm{d}r^{2}\right) +r^{2\left( 1-\sigma \right) }\mathrm{d}\varphi
^{2}+r^{2\sigma }l^{2}\mathrm{d}\alpha ^{2}~,  \label{LCcomp}
\end{equation}%
with $\sqrt{\mathfrak{g}}=r^{2\sigma ^{2}-2\sigma +1}l$, the Levi-Civita
spacetime being recovered in the $l\rightarrow \infty $ limit. A closed
spacelike $2$-surface $\mathcal{S}$ encompassing the singular axis (now a
ring) is given by constant $t$ and $r$, together with $0\leq \varphi \leq
2\pi $ and the compactified coordinate $0\leq \alpha \leq 2\pi $.

The Killing $1$-form $\boldsymbol{\xi }=-r^{2\sigma \left( \sigma -1\right) }%
\mathrm{d}t$ has the exterior derivative 
\begin{equation}
\mathrm{d}\boldsymbol{\xi }=2\sigma \left( \sigma -1\right) r^{2\sigma
^{2}-2\sigma -1}\mathrm{d}t\wedge \mathrm{d}r~,
\end{equation}%
with Hodge dual 
\begin{equation}
\ast \mathrm{d}\boldsymbol{\xi }=-2\sigma \left( \sigma -1\right) l\mathrm{d}%
\varphi \wedge \mathrm{d}\alpha ~.
\end{equation}%
Therefore the Komar mass emerges as 
\begin{equation}
m_{K}=\frac{\sigma \left( \sigma -1\right) l}{4\pi }\int_{0}^{2\pi }\!\!%
\mathrm{d}\varphi \int_{0}^{2\pi }\!\!\mathrm{d}\alpha =\pi \sigma \left(
\sigma -1\right) l~,
\end{equation}%
diverging for $l\rightarrow \infty $. Nevertheless the Komar mass density $%
\mu _{K}=m_{K}/2\pi l$ results in a finite constant 
\begin{equation}
\mu _{K}=\frac{\sigma \left( \sigma -1\right) }{2}=\frac{\lambda }{\left(
1-2\lambda \right) ^{2}}~,  \label{muK}
\end{equation}%
independent of the length scale $l$. Next, we take $l\rightarrow \infty $ to
obtain the original uncompactified spacetime, with $\mu _{K}$ unaffected by
this procedure.

Note than when $\lambda $ is small, $\mu _{K}\approx \lambda $ holds.

As the Komar mass, the Komar mass density also depends on the choice of the
timelike Killing vector. In the absence of a preferred normalization, it is
adapted to the temporal coordinate of the actual coordinate system. In
particular, the timelike coordinate vector in the Weyl form of the metric
[see the first Eq. (\ref{ERWtraf})] would give $\left( \mu _{K}\right) _{%
\mathrm{Weyl}}=\lambda $, another Komar mass density introduced in Ref. \cite%
{CostaNatarioSantos}.

Furthermore, as some of the coordinates may be accelerating, in principle $%
\mu _{K}$ could also include acceleration effects. We illustrate this point
in Appendix \ref{KomarR} for the Rindler metric.

\subsection{Levi-Civita spacetime parametrized by Komar mass density}

The Komar mass density becomes negative in the repulsive range $\sigma \in
\left( 0,1\right) $, with a minimal value of $-1/8$ at $\sigma =1/2$ ($%
\lambda =-1/2$). For all other values of $\sigma $ it stays positive, hence
its range is $\mu _{K}\geq -1/8$.

For both parameter values $\sigma =0,1$, where the metric is flat, $\mu _{K}$
vanishes. Hence, $\mu _{K}$ is better suited for the physical interpretation
of the spacetime than $\lambda $. The Rindler limit arises for $\mu
_{K}\rightarrow \infty $ (thus $\sigma \rightarrow \pm \infty $ or $\lambda
=1/2)$.

With the parameters $\sigma $ or $\lambda $ representing shorthand notations%
\begin{eqnarray}
\sigma &=&\frac{1\pm \sqrt{1+8\mu _{K}}}{2}~,  \notag \\
\lambda &=&\frac{1+4\mu _{K}\pm \sqrt{1+8\mu _{K}}}{8\mu _{K}}~,
\label{sila}
\end{eqnarray}%
cf. Eq. (\ref{muK}), the Levi-Civita metric is rewritten in terms of $\mu
_{K}$ as 
\begin{eqnarray}
\mathrm{d}s^{2} &=&r^{4\mu _{K}}\left( -\mathrm{d}t^{2}+\mathrm{d}%
r^{2}\right) +r^{1\mp \sqrt{1+8\mu _{K}}}\mathrm{d}\varphi ^{2}  \notag \\
&&+r^{1\pm \sqrt{1+8\mu _{K}}}\mathrm{d}z^{2}~.
\end{eqnarray}%
The duality in the ranges of $\lambda $ and $\sigma $, related to the
interchange of the coordinates $z$ and $\varphi $ is represented here by the
sign ambiguity, which selects the coordinate to be regarded as azimuthal
(periodic with $2\pi $) in flat space. Denoting this by $\psi \in \left[
0,2\pi \right] $ and the remaining axial coordinate by $Z$, the metric
becomes 
\begin{eqnarray}
\mathrm{d}s^{2} &=&r^{4\mu _{K}}\left( -\mathrm{d}t^{2}+\mathrm{d}%
r^{2}\right) +r^{1+\sqrt{1+8\mu _{K}}}\mathrm{d}\psi ^{2}  \notag \\
&&+r^{1-\sqrt{1+8\mu _{K}}}\mathrm{d}Z^{2}~.  \label{LCER}
\end{eqnarray}%
This choice is consistent with picking up the lower signs in Eq. (\ref{sila}%
). We will explore this parametrization of the Levi-Civita metric in what
follows.

\subsection{C-energy}

Thorne has proposed an energy-like quantity suitable for characterizing
systems with whole-cylinder symmetry. This is the cylindrical or C-energy 
\cite{Thorne}, arising as the projection of a covariantly conserved flux
vector to the worldline of the observer. For the Levi-Civita spacetime
outside a homogeneous cylinder the C-energy agrees with the mass per unit
length, but only when the latter is small and the pressures inside the
cylinder are negligible \cite{Thorne}.

\subsubsection{C-energy in terms of Komar mass density}

We adopt the recipe $E_{C}=\frac{1}{8}\ln \left( g_{rr}g_{ZZ}\right) $,
which is\thinspace $1/4$ of the C-energy as defined in Ref. \cite{Bini}
(based on the presentation of Chandrasekhar \cite{Chandrasekhar} of an
argument by Reula, which can be traced back to the integral of a suitable
Hamiltonian density in the radial direction). Our definition (after suitable
changes of notation) agrees with the one of Thorne \cite{Thorne}, and has
the correct Newtonian limit, as will be shown below. We obtain%
\begin{equation}
E_{C}=\frac{1+4\mu _{K}-\sqrt{1+8\mu _{K}}}{8}\ln r~,  \label{EC}
\end{equation}%
which increases monotonically with $\mu _{K}\geq 0$ and approximates $\mu
_{K}^{2}\ln r$ for small $\mu _{K}$.

When rewriting the above C-energy in terms of $\sigma $ or $\lambda $, the
monotonic increase holds for a positive $\sigma $ but it is lost for $%
\lambda $ (it holds only for small $\lambda $):%
\begin{equation}
E_{C}=\frac{\sigma ^{2}}{4}\ln r=\frac{\lambda ^{2}}{\left( 2\lambda
-1\right) ^{2}}\ln r~.  \label{EC1}
\end{equation}%
This supports the naturalness of the parametrization of the Levi-Civita
metric with the Komar mass density calculated in Einstein-Rosen coordinates,
rather than in terms of $\left( \mu _{K}\right) _{\mathrm{Weyl}}=\lambda $.

\subsubsection{Alternative definition of C energy}

{Another definition of C-energy, 
\begin{equation}
E_{C}^{\mathrm{alt}}=\frac{1}{8}\left( 1-\frac{1}{g_{rr}g_{ZZ}}\right) 
\end{equation}%
has been added in the proof of Thorne's paper \cite{Thorne}, with the
purpose of all observers measuring finite C-energy density under all
circumstances. The weak gravity limit (given by $g_{rr}\approx g_{ZZ}\approx
1$) is the same for both definitions of C energy. }This alternative
definition is explored in the works of Hayward \cite{Hayward} and Chiba%
\textbf{\ }\cite{Chiba}, which claims that $E_{C}^{\mathrm{alt}}$ arises as
the integral of a suitable Hamiltonian with reference to Chandrasekhar's
work \cite{Chandrasekhar}, however this rather leads to $E_{C}$.

Note that for the Levi-Civita spacetime, $E_{C}$ is also finite everywhere
apart from the singularity.

\subsubsection{Newtonian limit}

A Lagrangian density leading to the Poisson equation $\nabla ^{2}\phi
_{N}=4\pi \rho $ (with $\phi _{N}$ the Newtonian gravitational potential and 
$\rho $ the possibly distributional mass density) is%
\begin{equation}
\mathcal{L}=-\frac{1}{8\pi }\left( \mathbf{\nabla }\phi _{N}\right)
^{2}-\rho \phi _{N}~.
\end{equation}%
This consists entirely of the potential term of gravity and an interaction
contribution. Hence the volume density of the Newtonian gravitational energy 
\begin{equation}
\mathcal{E}=\frac{1}{8\pi }\left( \mathbf{\nabla }\phi _{N}\right) ^{2}
\end{equation}%
for $\phi _{N}=2\lambda \ln r$ becomes $\mathcal{E}=\lambda ^{2}/\left( 2\pi
r^{2}\right) $. Integrating this between two cylindrical surfaces and taking
its density $E_{N}$ along the $Z$ axis yields%
\begin{equation}
E_{N}\left( r_{2}\right) -E_{N}\left( r_{1}\right) =\lambda ^{2}\ln
r_{2}-\lambda ^{2}\ln r_{1}~.
\end{equation}%
This agrees with the Newtonian limit of the difference $E_{C}\left(
r_{2}\right) -E_{C}\left( r_{1}\right) $ of the C-energies.

\section{Geometric characterization in terms of the Komar mass density}

\subsection{Curvature invariants}

\begin{figure}[th]
\includegraphics[scale=0.25]{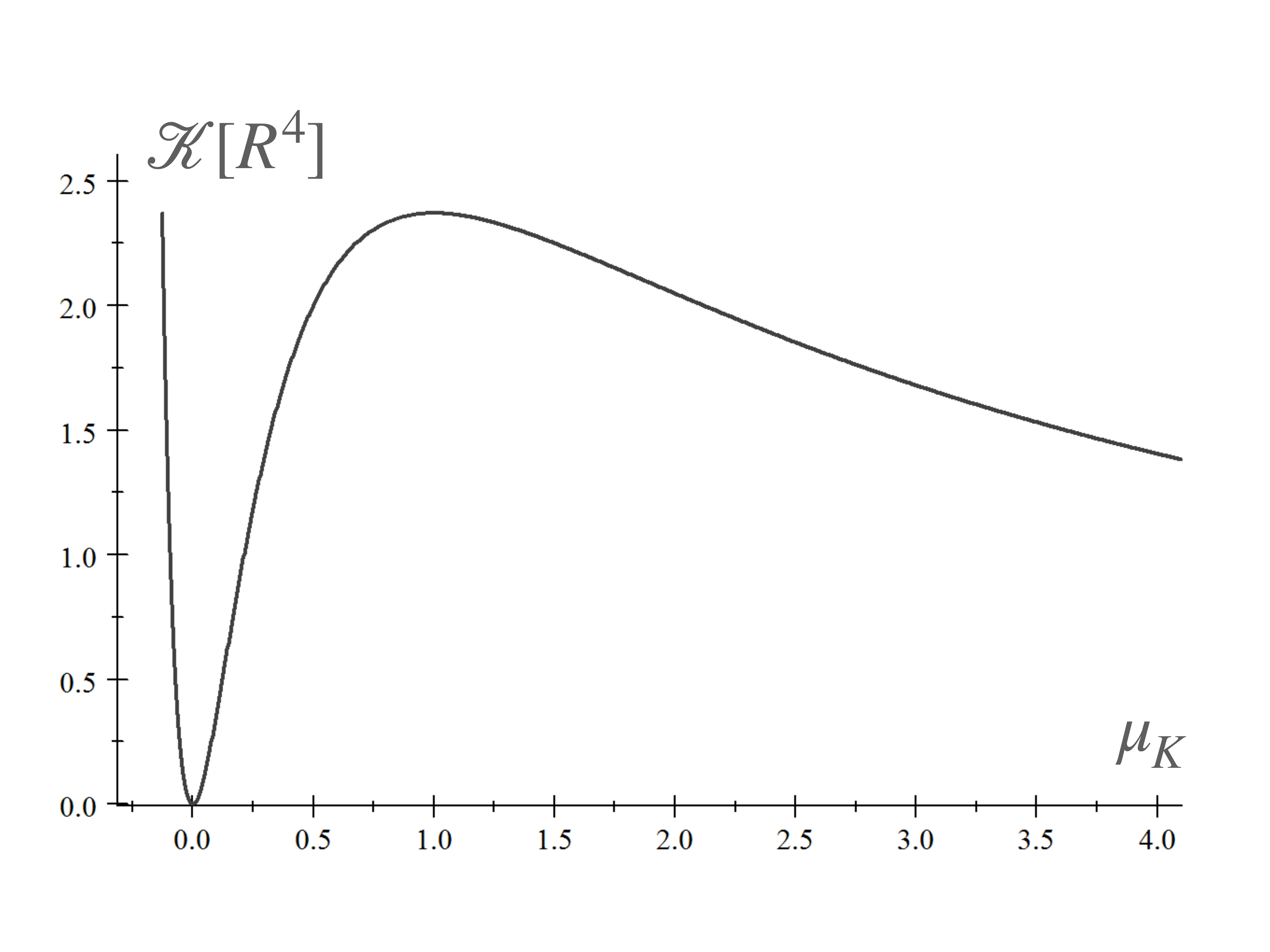}
\caption{The Kretschmann scalar as function of the Komar mass density (in units of $R^{4}$)}
\label{KretschFig}
\end{figure}

In order to characterize the radial features of the spacetime, we introduce
the proper radial distance 
\begin{equation}
R=\int_{0}^{r}r^{2\mu _{K}}\mathrm{d}r=\frac{r^{1+2\mu _{K}}}{1+2\mu _{K}}~.
\label{proprad}
\end{equation}%
The Kretschmann scalar 
\begin{equation}
\mathcal{K}=\frac{64\mu _{K}^{2}}{\left( 1+2\mu _{K}\right) ^{3}}R^{-4}~
\end{equation}%
scales with $\mu _{K}$, confirming a flat spacetime when it vanishes. It
also vanishes in the limit $\mu _{K}\rightarrow \infty $. For all other
parameter values the Kretschmann scalar falls off at $R\rightarrow \infty $
and there is a naked singularity on the $Z$ axis.

\begin{figure}
\includegraphics[scale=0.23]{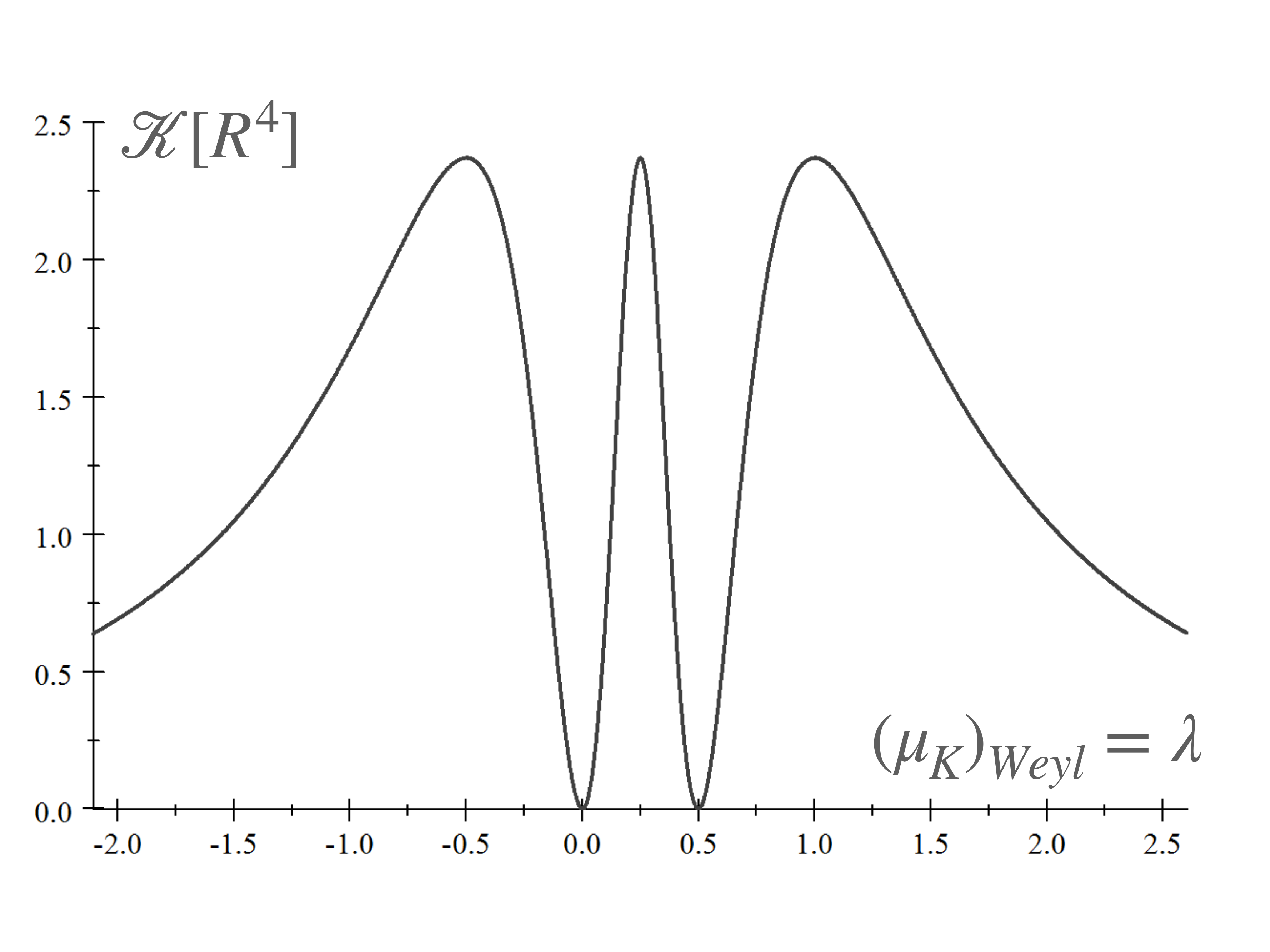}
\caption{The Kretschmann scalar as function of the Komar mass density defined with the Weyl time (in units of $R^{4}$) shows a double coverage as compared to the Komar mass density defined with the Einstein-Rosen time.}
\label{KretschWFig}
\end{figure}

The dependence of the Kretschmann scalar on the Komar mass density is
illustrated for $R=1$ on Fig. \ref{KretschFig}. It has three extrema: (i) it
vanishes at $\mu _{K}=0$, as the spacetime becomes flat; (ii) at $\mu
_{K}=-1/8$ there is the maximum in the negative $\mu _{K}$ range; (iii) at $%
\mu _{K1}=1$ the positive $\mu _{K}$ range has maximal Kretschmann
curvature. The existence of $\lambda _{1}=1/4$ (corresponding to $\mu _{K1}$%
) as the value of the parameter generating maximal Kretschmann curvature has
been emphasized by Bonnor and Martins in their discussion of the Levi-Civita
metric in the Weyl form \cite{Bonnorlambda4} (see Fig. \ref{KretschWFig} for
the Kretschmann scalar as function of $\left( \mu _{K}\right) _{\mathrm{Weyl}%
}=\lambda $ for $R=1,$ showing a double degeneracy, when compared to Fig. %
\ref{KretschFig}). The increase and subsequent decrease of the Kretschmann
curvature with $\mu _{K}$ is counterintuitive, undermining the
interpretation of $\mu _{K}$ as mass density

\begin{figure}
\includegraphics[scale=0.23]{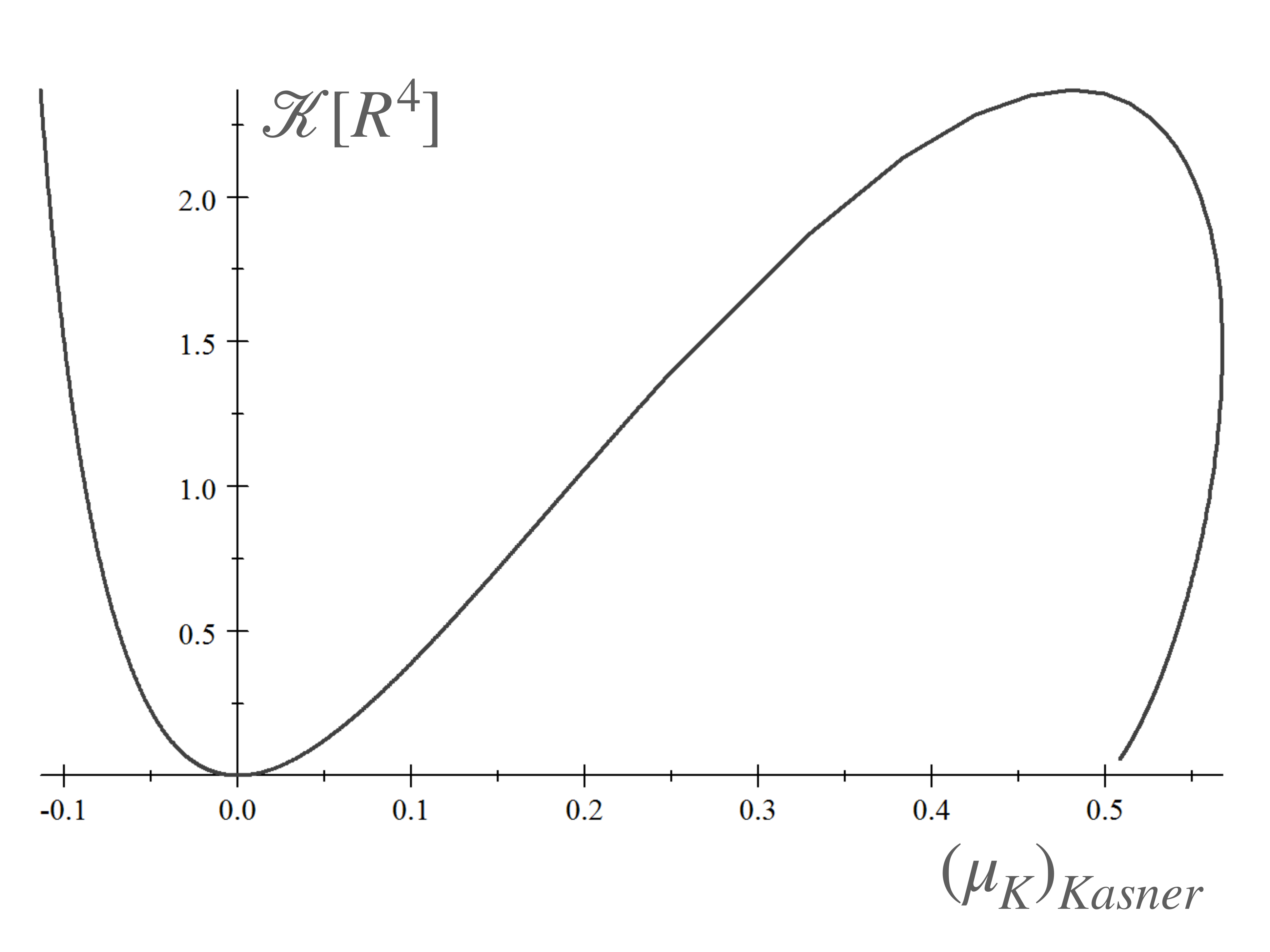}
\caption{The Kretschmann scalar as function of the Komar mass density defined with the Kasner time (in units of $R^{4}$) is a double valued function for $\left( \protect\mu _{K}\right) _{\mathrm{Kasner}}>0.5$. By contrast $\mathcal{K}$ is well defined everywhere in terms of the Komar mass density $\protect\mu _{K}$ defined with the Einstein-Rosen time}
\label{KretschKasner}
\end{figure}
The curvature of a vacuum spacetime, consisting purely of the Weyl tensor $%
C_{ijkl}$ is characterized by four scalar invariants \cite{Harvey}:%
\begin{eqnarray}
J_{1} &=&A_{ij}^{~\ ij}~,\quad J_{2}=B_{ij}^{~\ ij}~,  \notag \\
J_{3} &=&A_{ij}^{~\ kl}A_{kl}^{~\ ij}-\frac{J_{1}^{2}}{2}~,  \notag \\
J_{4} &=&A_{ij}^{~\ kl}B_{kl}^{~\ ij}-\frac{5J_{1}J_{2}}{12}~,
\end{eqnarray}%
with%
\begin{equation}
A_{ijkl}=C_{ij}^{~\ mn}C_{mnkl}~,\quad B_{ijkl}=C_{ij}^{~\ mn}A_{mnkl}~.
\end{equation}%
For the Levi-Civita metric (\ref{LCER}) the scalars are 
\begin{eqnarray}
J_{1} &=&\frac{64\mu _{K}^{2}\left( 1+2\mu _{K}\right) }{r^{4\left( 1+2\mu
_{K}\right) }}=\mathcal{K~},  \notag \\
J_{2} &=&-\frac{768\mu _{K}^{4}}{r^{6\left( 1+2\mu _{K}\right) }}=-\frac{%
3\mu _{K}}{2}\left( \frac{\mathcal{K}}{1+2\mu _{K}}\right) ^{3/2}~,  \notag
\\
J_{3} &=&-\frac{1024\mu _{K}^{4}\left( 1+2\mu _{K}\right) ^{2}}{r^{8\left(
1+2\mu _{K}\right) }}=-\frac{\mathcal{K}^{2}}{4}~,  \notag \\
J_{4} &=&0~.  \label{scalars}
\end{eqnarray}%
Hence all invariants are expressed in terms of the Kretschmann scalar and
exhibit similar dependence on $\mu _{K}$.

\subsection{Kasner form and Rindler limit of the Levi-Civita spacetime}

As in the limit $\mu _{K}\rightarrow \infty $ the metric (\ref{LCER})
diverges, we rewrite it in a more suitable form in terms of the proper
radial distance (\ref{proprad}) and the rescaled time%
\begin{equation}
T=\left( 1+2\mu _{K}\right) ^{\frac{2\mu _{K}}{1+2\mu _{K}}}t~,
\end{equation}%
as new coordinates. The line element becomes%
\begin{eqnarray}
\mathrm{d}s^{2} &=&-R^{2p_{0}}\mathrm{d}T^{2}+\mathrm{d}R^{2}+\left[ \left(
1+2\mu _{K}\right) R\right] ^{2p_{+}}\mathrm{d}\psi ^{2}  \notag \\
&&+\left[ \left( 1+2\mu _{K}\right) R\right] ^{2p_{-}}\mathrm{d}Z^{2}~,
\label{LCfinal}
\end{eqnarray}%
or 
\begin{equation}
\mathrm{d}s^{2}=-R^{2p_{0}}\mathrm{d}T^{2}+\mathrm{d}R^{2}+R^{2p_{+}}\mathrm{%
d}\chi ^{2}+R^{2p_{-}}\mathrm{d}\zeta ^{2}~,  \label{LCKasner}
\end{equation}%
with%
\begin{eqnarray}
\chi &=&\left( 1+2\mu _{K}\right) ^{p_{+}}\psi ~,  \notag \\
\zeta &=&\left( 1+2\mu _{K}\right) ^{p_{-}}Z~,
\end{eqnarray}%
and the coefficients%
\begin{equation}
p_{0}=\frac{2\mu _{K}}{1+2\mu _{K}}~,\quad p_{\pm }=\frac{1\pm \sqrt{1+8\mu
_{K}}}{2\left( 1+2\mu _{K}\right) }~.
\end{equation}%
The ranges of the coordinates are $T,\zeta \in \mathbb{R}$, $R\in \mathbb{R}%
^{+}$ and $\chi \in \left[ 0,2\pi \left( 1+2\mu _{K}\right) ^{p_{+}}\right] $%
. The powers obey $p_{0}+p${}$_{+}+p_{-}=1$ and $p_{0}^{2}+p${}$%
_{+}^{2}+p_{-}^{2}=1$, implying $p_{0}\in \left[ -1/3,1\right] $, $p_{+}\in %
\left[ 0,1\right] $, and $p_{-}\in \left[ -1/3,2/3\right] $. This form of
the Levi-Civita metric resembles the inhomogeneous Kasner metric with
coordinates $T$ and $R$ interchanged \cite{GriffithsP} and is expressed in
terms of the Komar mass per unit $Z$ of the Einstein-Rosen coordinates.

For the Levi-Civita metric when $\mu _{K}\rightarrow \infty $, the
coefficients reduce to $p_{0}\rightarrow 1$, $p_{\pm }\rightarrow 0$ and $%
\left( 1+2\mu _{K}\right) ^{2p_{\pm }}\rightarrow 1$, hence 
\begin{equation}
\mathrm{d}s_{\mu _{K}\rightarrow \infty }^{2}=-R^{2}\mathrm{d}T^{2}+\mathrm{d%
}R^{2}+\mathrm{d}Z^{2}+\mathrm{d}\psi ^{2}~.  \label{Rindler}
\end{equation}%
In this limit it simplifies to a Rindler metric with particular topology $%
S^{1}\times \mathbb{R}^{3}$, representing flat spacetime perceived by a
uniformly accelerated observer with acceleration $R^{-1}$ along $R$. The
unusual topology consists of each point of the Rindler wedge $\left(
T,R\right) $ corresponding to a cylinder of unit radius (parametrized by
longitudinal and angular variables $Z$ and $\psi $, respectively), rather
than a plane.

In the $\left( T,R\right) $ coordinates the Komar mass density (associated
to the time coordinate vector) emerges as 
\begin{equation}
\left( \mu _{K}\right) _{\mathrm{Kasner}}=\frac{p_{0}}{2}\left( 1+2\mu
_{K}\right) ^{p_{-}+p_{+}}=\frac{\mu _{K}}{\left( 1+2\mu _{K}\right) ^{\frac{%
2\mu _{K}}{1+2\mu _{K}}}}~.  \label{KomarTR}
\end{equation}%
In the Rindler limit $\mu _{K}\rightarrow \infty $ of an accelerated
observer in flat spacetime $\left( \mu _{K}\right) _{\mathrm{Kasner}}=1/2$.
This is consistent with the Komar mass surface density $\sigma _{K}$ of the
Rindler spacetime given by Eq. (\ref{sigmaK}), multiplied by the
circumference $2\pi $ of the additional compactified coordinate.

We have already seen that $\mu _{K}$ bears the advantage over $\left( \mu
_{K}\right) _{\mathrm{Weyl}}=\lambda $ of avoiding a double coverage of the
parameter space. The parameter $\left( \mu _{K}\right) _{\mathrm{Kasner}}$
suffers from another inconvenience, the Kretschmann scalar $\mathcal{K}$
turning out as a multivalued function of $\left( \mu _{K}\right) _{\mathrm{%
Kasner}}$ in the range $\left( \mu _{K}\right) _{\mathrm{Kasner}}\geq 1/2$
(which applies to all $\mu _{K}\geq 1.1466$), as can be seen from Fig. \ref%
{KretschKasner}. Hence we keep $\mu _{K}$ for parametrizing the metric.

With increasing $\mu _{K}$ the Levi-Civita metric approaches the Rindler
limit, supporting the statement that beside mass and gravitational energy,
the Komar mass density $\mu _{K}$ also encompasses acceleration effects. As
the Rindler observers (from the point of view of an inertial observer)
accelerate as $R^{-1}$ in the $R$ direction, the Rindler spacetime can also
be interpreted through the equivalence principle as a (Newtonian)
gravitational field homogeneous in the $T$, $Z$, and $\psi $ directions
(hence with the coordinate lines $R$ becoming parallel with $\mu
_{K}\rightarrow \infty $). Thus, we conjecture that the magnitude of $\mu
_{K}$\ correlates with the degree of homogeneity (as defined above) of the
Newtonian gravitational field.

\subsection{The singular axis\label{Sing}}

Recently, Ref. \cite{Ahmed} has presented the cylindrically symmetric vacuum
spacetime%
\begin{eqnarray}
\mathrm{d}s^{2}\! &=&\!\sinh ^{2}\!r_{\ast }\!\left( -\mathrm{d}t_{\ast
}^{2}\!+\!\mathrm{d}\varphi _{\ast }^{2}\right) +\frac{\mathrm{d}z_{\ast
}^{2}}{\sinh r_{\ast }}  \notag \\
&&+\cosh ^{2}\!r_{\ast }\sinh r_{\ast }\mathrm{d}r_{\ast }^{2}~,
\label{Farook}
\end{eqnarray}%
claiming, among others, that the $r_{\ast }=0$ axis is a geodesically
incomplete (for null geodesics) soft singularity both in the sense of Królak 
\cite{Krolak} and of Tipler \cite{Tipler}. However, we show in Appendix \ref%
{AhmedNO} that this metric is but a particular case of the Levi-Civita
metric, corresponding to $\mu _{K}=1$, thus referring to the value of the
parameter, where the Kretschmann curvature is maximal. We also correct the
curvature invariants given in Ref. \cite{Ahmed} and find that the authors of
Ref. \cite{Ahmed} incorrectly applied the strong singularity criteria.

Therefore we present in this section the rigorous analysis of the
singularity on the symmetry axis of the Levi-Civita spacetime, for a generic
value of $\mu _{K}$.

Any strong singularity crushes to zero all $3$-volumes (or $2$-volumes,
respectively) parallel transported along timelike (or null) geodesics \cite%
{EllisSchmidt}. The concept was formulated rigorously by Tipler \cite{Tipler}%
. Based on the expectation that in physically realistic spacetimes
singularities are both strong and hidden by horizons, Królak proposed a less
restrictive condition on the convergence of geodesics {\cite{Krolak}}.
Clarke and Królak {\cite{ClarkeKrolak}} formulated computational recipes
corresponding to either the necessary or the sufficient conditions for the
Tipler and Królak criteria. For timelike geodesics, there are no conditions
that are simulataneously necessary and sufficient, therefore we focus on
null geodesics, for which (with special conditions holding on the Weyl
tensor), necessary and sufficient conditions may coincide.

In particular, if the Weyl tensor is not identically zero, and it does not
display oscillatory behaviour along a null geodesic $\gamma :I\subseteq 
\mathbb{R}\rightarrow M$ hitting the singularity at affine parameter $%
\lambda \rightarrow \lambda _{s}$, then the Królak strong singularity
condition is satisfied if and only if any component of 
\begin{equation}
N_{\ b}^{a}\left( \lambda \right) =\int_{0}^{\lambda }\mathrm{d}\lambda
^{\prime }\left( \int_{0}^{\lambda ^{\prime }}\mathrm{d}\lambda ^{\prime
\prime }\left\vert C_{\ 0b0}^{a}\left( \lambda ^{\prime \prime }\right)
\right\vert \right) ^{2}  \label{NKro}
\end{equation}%
diverge as $\lambda \rightarrow \lambda _{s}$, while the Tipler strong
singularity condition holds if and only if any component of 
\begin{equation}
L_{\ b}^{a}\left( \lambda \right) =\int_{0}^{\lambda }\mathrm{d}\lambda
^{\prime }\int_{0}^{\lambda ^{\prime }}\mathrm{d}\lambda ^{\prime \prime
}\left( \int_{0}^{\lambda ^{\prime \prime }}\mathrm{d}\lambda ^{\prime
\prime \prime }\left\vert C_{\ 0b0}^{a}\left( \lambda ^{\prime \prime \prime
}\right) \right\vert \right) ^{2}  \label{LTip}
\end{equation}%
diverge as $\lambda \rightarrow \lambda _{s}$. The components $C_{\ bcd}^{a}$
are calculated with respect to a pseudo-orthonormal frame $e_{\mathbf{0}},e_{%
\mathbf{1}},e_{\mathbf{2}},e_{\mathbf{3}}$ parallel propagated along the
geodesic, such that $e_{\mathbf{0}}$ and $e_{\mathbf{1}}$ are the null
vectors, $e_{\mathbf{2}}$ and $e_{\mathbf{3}}$ are the spacelike vectors and 
$e_{\mathbf{0}}=\dot{\gamma}$ is the geodesic's tangent.

We consider radial null geodesics in the $\left( T,R\right) $-plane of the
Kasner-like coordinates: 
\begin{equation}
\gamma \left( \lambda \right) =\left( T\left( \lambda \right) ,R\left(
\lambda \right) ,\chi _{0},\zeta _{0}\right) ~,
\end{equation}%
where the latter two components are constants. The velocity vector is 
\begin{equation}
e_{\mathbf{0}}\left( \lambda \right) =\left( \dot{T}\left( \lambda \right) ,%
\dot{R}\left( \lambda \right) ,0,0\right) ~,
\end{equation}%
where the overdot denotes derivative with respect to the affine parameter. A
radially ingoing null geodesic satisfies the equation 
\begin{equation}
\dot{T}=R^{-2p_{0}}~,\quad \dot{R}=-R^{-p_{0}}~,
\end{equation}%
with an irrelevant constant of integration set to unity. Explicitly
integrating the geodesic equations is possible, but unnecessary, as the
integrals (\ref{NKro}) and (\ref{LTip}) can be calculated through the chain
rule as 
\begin{equation}
\int_{0}^{\lambda _{s}}\mathrm{d}\lambda =\int_{0}^{R_{0}}R^{p_{0}\,}\mathrm{%
d}R\,~,
\end{equation}%
where $R_{0}=R\left( 0\right) $. A parallel frame along the radial null
geodesic is then given by 
\begin{align}
e_{\mathbf{0}}& =\left( R^{-2p_{0}},-R^{-p_{0}},0,0\right) ~,  \notag \\
e_{\mathbf{1}}& =\left( \frac{1}{2},\frac{1}{2}R^{p_{0}},0,0\right) ~, 
\notag \\
e_{\mathbf{2}}& =\left( 0,0,R^{-p_{+}},0\right) ~,  \notag \\
e_{\mathbf{3}}& =\left( 0,0,0,R^{-p_{-}}\right) ~.  \label{e03}
\end{align}%
The Weyl tensor, also the integrals (\ref{NKro}) and (\ref{LTip}) are
calculated in the above frame in Appendix \ref{StrongSing}. We found that
the components $L_{\ 2}^{2}$ and $L_{\ 3}^{3}$ of the Tipler integrals, as
well as the components $N_{\ 2}^{2}$ and $N_{\ 3}^{3}$ of the Królak
integrals diverge logarithmically with $R\rightarrow 0$, provided $\mu
_{K}\neq 0,\infty $. Hence radial null geodesics satisfy both the Tipler and
Królak singularity conditions.

In conclusion, the symmetry axis of the Levi-Civita spacetime represents a
strong curvature singularity for any of the allowed parameter values
(including $\mu _{K}=1$, which refutes the claim made in Ref. {\cite{Ahmed}}
about the singularity being soft).

\section{Physical characterization in terms of the Komar mass density}

In this section we analyze the the gravitational effects ocurring in the
Levi-Civita spacetime, both from a Newtonian and a general relativistic
perspective.

\subsection{Gravitational acceleration}

In what follows, we investigate the Levi-Civita spacetime by considering a
stationary observer at fixed proper distance from the $Z$-axis, with
4-velocity $u^{a}=\left( R^{-p_{0}},0,0,0\right) $ and 4-acceleration 
\begin{equation}
a^{a}\equiv u^{b}\nabla _{b}u^{a}=\frac{2\mu _{K}}{1+2\mu _{K}}R^{-1}\delta
_{R}^{a}~,
\end{equation}%
the latter compensating for the gravitational effect of the cylinder,
according to the equivalence principle. The acceleration changing sign with $%
\mu _{K}$ shows that gravity is repulsive for $-1/8\leq \mu _{K}<0$ and
attractive for $\mu _{K}>0$. In the latter case it approximates the
Newtonian regime $-d\phi _{N}/dr$ at small $\mu _{K}$ (with $\phi _{N}$ the
Newtonian potential generated by a linear mass distribution on the $Z$ axis)
and decays at $R\rightarrow \infty $, as expected.

We define the gravitational acceleration (in a Newtonian sense) 
\begin{equation}
a_{g}=-\frac{2\mu _{K}}{1+2\mu _{K}}R^{-1}~,  \label{accG}
\end{equation}%
the magnitude of which represents the magnitude of the acceleration to keep
the observer in orbit and its sign being negative (positive) in the
attractive (repulsive) regime. This arises from the potential%
\begin{equation}
\phi =\frac{2\mu _{K}}{1+2\mu _{K}}\ln R~,
\end{equation}%
which reduces to the Newtonian limit $\phi _{N}=2\mu _{K}\ln r$ for small $%
\mu _{K}$.

\begin{figure}
\includegraphics[scale=0.25]{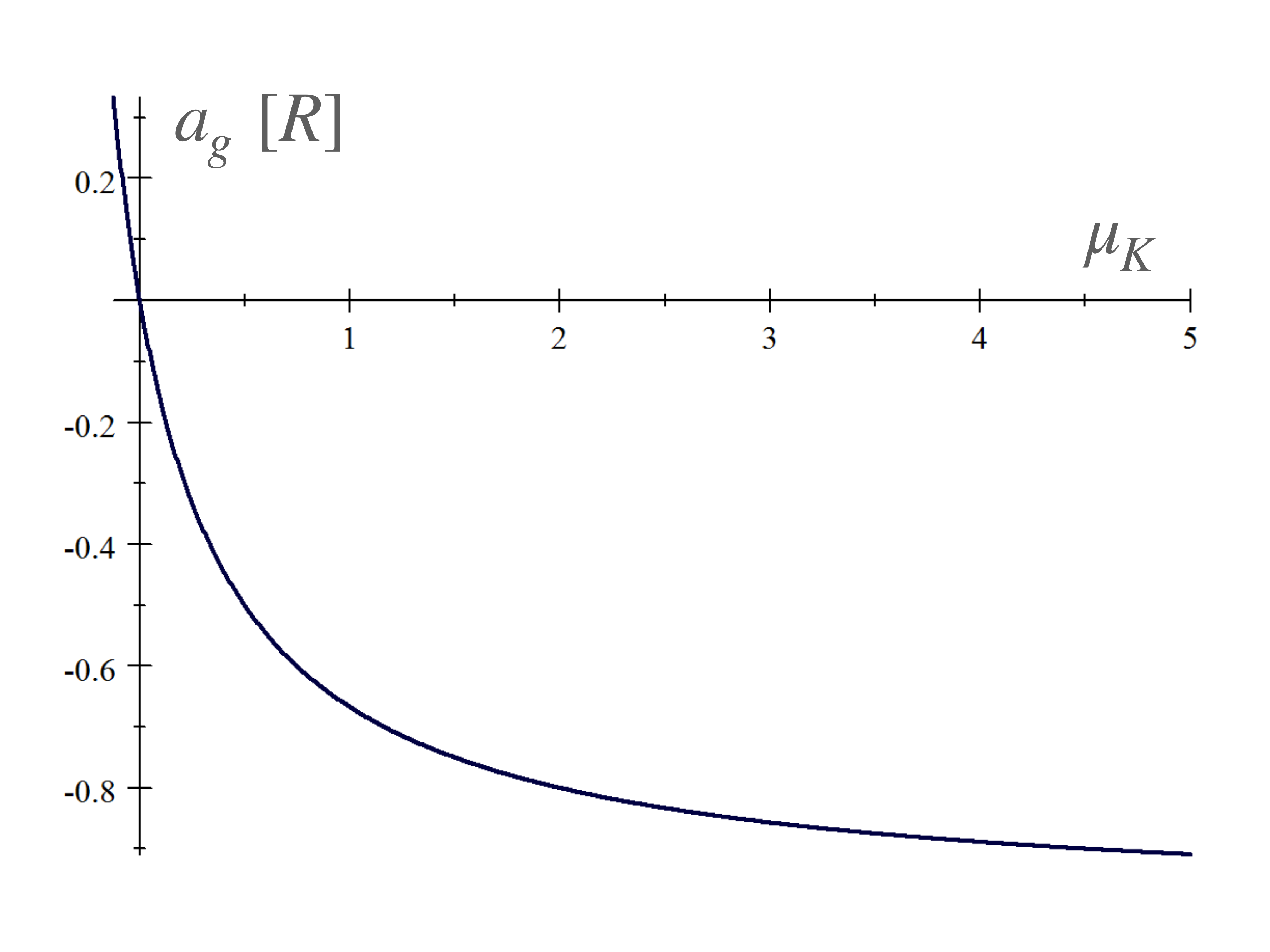}
\caption{The gravitational acceleration (defined in a Newtonian sense through the equivalence principle) as function of the Komar mass density, in units of $R$.}
\label{accFig}
\end{figure}

At unit proper radial distance $a_{g}$ becomes $a_{g}^{R=1}=-2\mu
_{K}/\left( 1+2\mu _{K}\right) $, illustrated on Fig. \ref{accFig}. The
gravitational acceleration at unit proper radial distance is negative
(positive) for positive (negative) $\mu _{K}$. For positive $\mu _{K}$ the
gravitational acceleration is attractive and increases monotonically with $%
\mu _{K}$, asymptoting $-1$. Hence, despite increasing $\mu _{K}$,
gravitational attraction cannot increase above a certain limit.

\subsection{Geodesic deviation}

We can better understand the gravitational field by discussing the geodesic
deviation. We consider the congruence $U^{a}\equiv \left( \partial /\partial
\tau \right) ^{a}$, with the proper time $\tau $ given by $d\tau
=R^{p_{0}}dT $, implying $U^{a}=R^{-p_{0}}\delta _{T}^{a}$ for any given $R$%
. Infinitesimally close curves of the congruence are separated by the
deviation vector $X^{a}\equiv \left( \partial /\partial R\right) ^{a}=\delta
_{R}^{a}$. By construction both $U^{a}$ and $X^{a}$ are normalized and they
commute. In any arbitrarily chosen point ($\tau =\tau _{1}$, $R=R_{1}$)
there is a geodesic with tangent $V^{a}\left( \tau _{1},R_{1}\right) \equiv
U^{a}\left( \tau _{1},R_{1}\right) $. In this point the acceleration $%
D^{2}X^{a}/d\tau ^{2}$ can be computed as 
\begin{eqnarray}
A^{a} &=&R_{~bcd}^{a}U^{b}U^{c}X^{d}=R^{-2p_{0}}R_{~TTR}^{a}  \notag \\
&=&p_{0}\left( 1-p_{0}\right) R^{-2}\delta _{R}^{a}~,
\end{eqnarray}%
with magnitude%
\begin{equation}
\left( g_{ab}A^{a}A^{b}\right) ^{1/2}=\frac{2\left\vert \mu _{K}\right\vert 
}{\left( 1+2\mu _{K}\right) ^{2}}R^{-2}~.  \label{accDev}
\end{equation}

\begin{figure}
\includegraphics[scale=0.23]{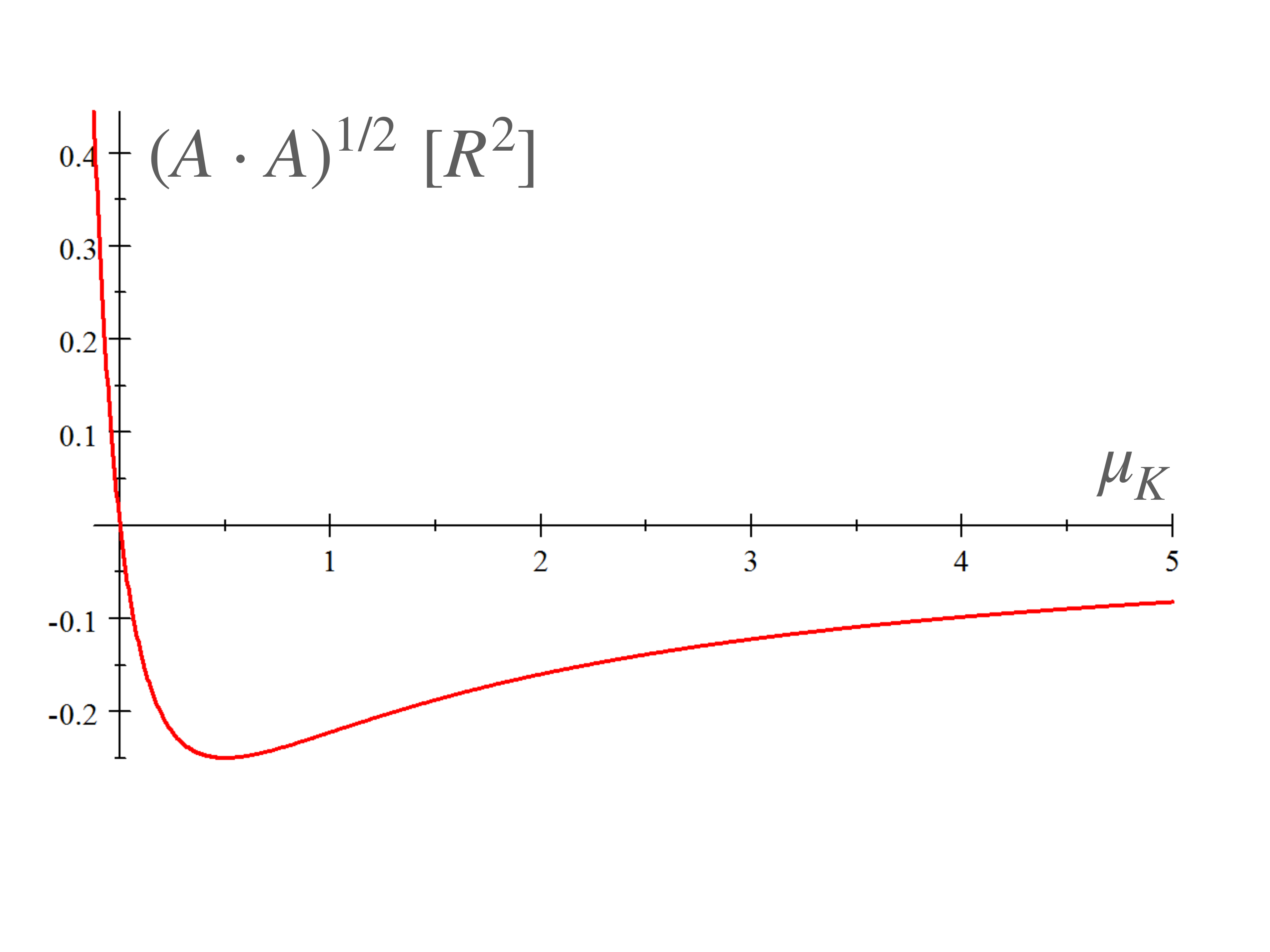}
\caption{The tidal acceleration among geodesic observers as function of the Komar mass density, in units of $R^{2}$.}
\label{geodDev}
\end{figure}

The negative of this acceleration represents the tidal force acting on a
unit mass particle. We visualize its dependence of $\mu _{K}$ at unit proper
radial distance on Fig. \ref{geodDev}

The tidal acceleration is zero at $\mu _{K}=0$, as expected for a\ flat
spacetime. Then it increases with $\mu _{K}$ (as for small values it
represents the mass density of the cylindric source) up to $\mu _{K2}=1/2$
(corresponding to $\sigma _{2}=\left( 1-\sqrt{5}\right) /2$ and $\lambda
_{2}=\left( 3-\sqrt{5}\right) /4$). Then it decreases again, falling off to
zero at $\mu _{K}\rightarrow \infty $. As we discussed before, this renders
the Levi-Civita spacetime into Rindler spacetime. In this limit the tidal
force vanishes and the gravitational field (in a Newtonian sense) attains a
high degree of homogeneity in the $T$, $Z$, and $\psi $ directions.

\section{Discussion and concluding remarks}

Levi-Civita spacetime can be regarded as the strong field background for a
cylindrical gravitational wave, which is the best testbed for comparing
quantization methods of gravitational waves. Hence it is paramount to
clearly understand this static spacetime. Previous presentations relied on
either of the parameters $\lambda $ or $\sigma $, emerging in the Weyl- or
Einstein-Rosen coordinates, respectively. Although both parameters, whenever
they are small and positive, have the nice interpretation of mass density
along the symmetry axis, when considered across their full allowed ranges,
are hard to interpret. The reason for this is twofold. First, there is a
double coverage of the parameter space, corresponding to a possible
interchange of the roles of the axial and polar variables. Then there are
two kinds of flat limits, the second one being of Rindler type. This leads
to a quartett of parameter values, all leading to flat limit.

Despite the axis being infinite, we could compute the Komar mass density $%
\mu _{K}$ along the axis through a compactification and a subsequent blowing
up of the compactification radius. By introducing $\mu _{K}$ as a new metric
parameter, we got rid of the double coverage, but the flat limit still
arises in two cases, for $\mu _{K}=0$ and $\mu _{K}\rightarrow \infty $. The
first indeed represents no mass on the axis. The second one is a Rindler
spacetime, as can be seen manifestly in Kasner type coordinates, which
include $R$, the proper radial distance measured from the axis.

The Komar mass density can be in the narrow negative range $-1/8\leq \mu
_{K}<0$, when gravity is repulsive. For all positive values it is
attractive. In the process of increasing $\mu _{K}$ from $0$ to $\infty $
the Kretschmann scalar increases to $1$, then it decreases again. We
identified a recently published solution in Ref. \cite{Ahmed} as the
Levi-Civita spacetime pertinent to the maximal Kretschmann scalar and
corrected a number of its claims. In the process we proved that the
singularity on the axis is strong for null geodesics both in the Tipler and
Królak senses, for any $\mu _{K}$.

In order to understand the Rindler limit at $\mu _{K}\rightarrow \infty $,
however $R$ cannot be regarded as radial any more. Initially $\left( R,\psi
\right) $ cover $\mathbb{R}^{2}$ as polar coordinates, while in the Rindler
limit they cover a cylinder $S^{1}\times \mathbb{R}$. This process can be
visualized (through an embedding into higher dimensions) as a pinching of
the $\left( R,\psi \right) $ plane into a direction perpendicular to both
the plane and the $Z$ axis, creating a cone-like shape. As $\mu _{K}$
further increases, the tip of the cone opens up into a funnel, which attains
its climax as a constant cross-section tube, in which the $R$ coordinate
lines become parallel. In the Rindler limit inertial observers have
coordinate acceleration $R^{-1}$ in the negative $R$ direction, appearing as
gravity (in the Newtonian sense), with the tidal acceleration vanishing.

\begin{figure}
\includegraphics[scale=0.23]{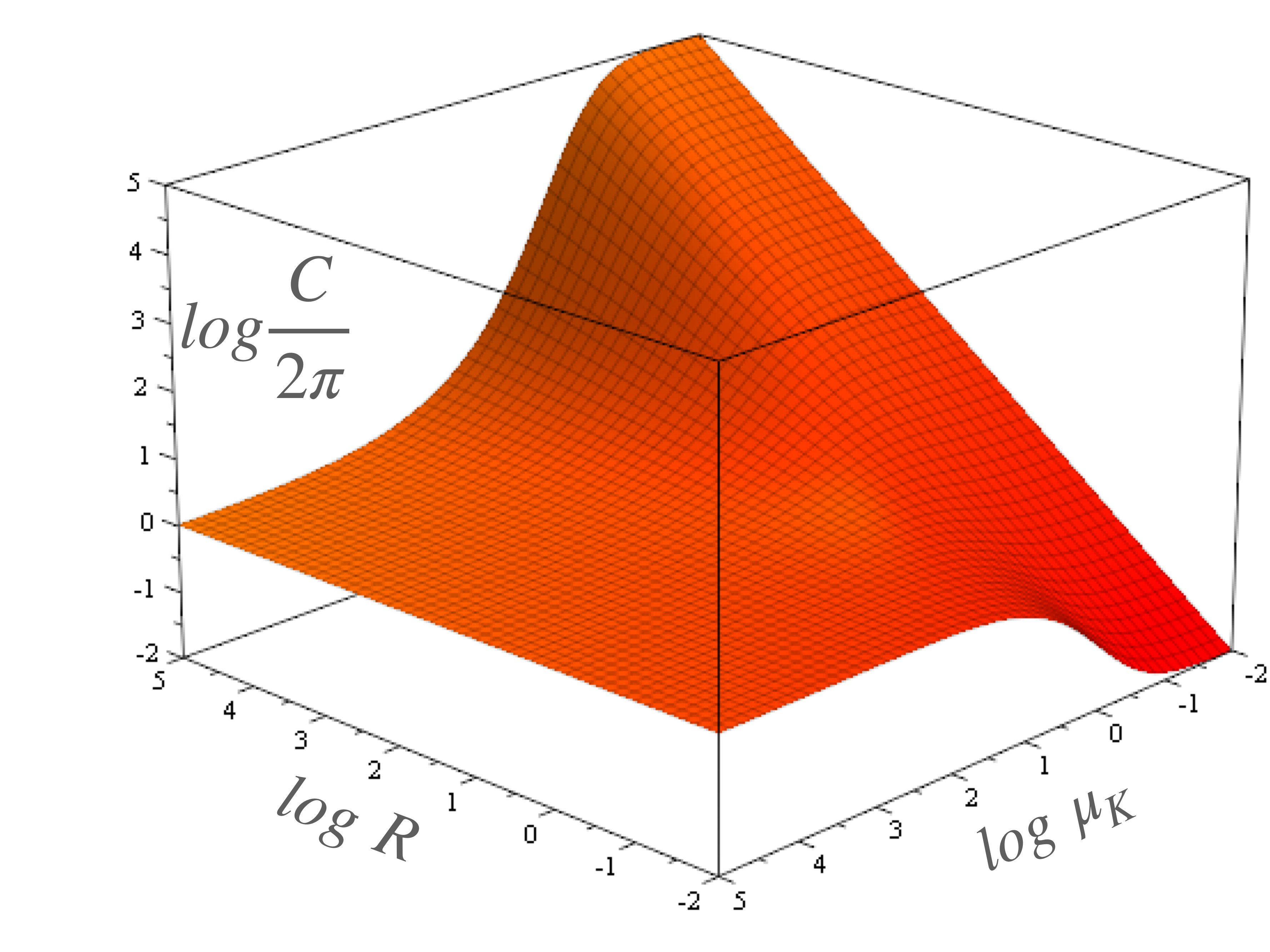}
\caption{The transformation of the geometry induced by an increasing Komar mass density $\protect\mu _{K}$. The circumference $C$ of a circle is approximately $2\protect\pi $ times the radius $R$ for small $\protect\mu _{K}$, as shown on the right side of the plotted surface, representing an almost flat, cylindrically symmetric spacetime. With increasing $\protect\mu _{K}$ the increase of the cicumference with radius becomes slower, eventually the circumference becoming a constant, regardless of the value of the coordinate $R$. The latter limit corresponds to the Rindler spacetime with poinwise acceleration $R^{-1}$ in the $R$ direction.}
\label{circumference3D}
\end{figure}

\begin{figure}
\includegraphics[scale=0.3]{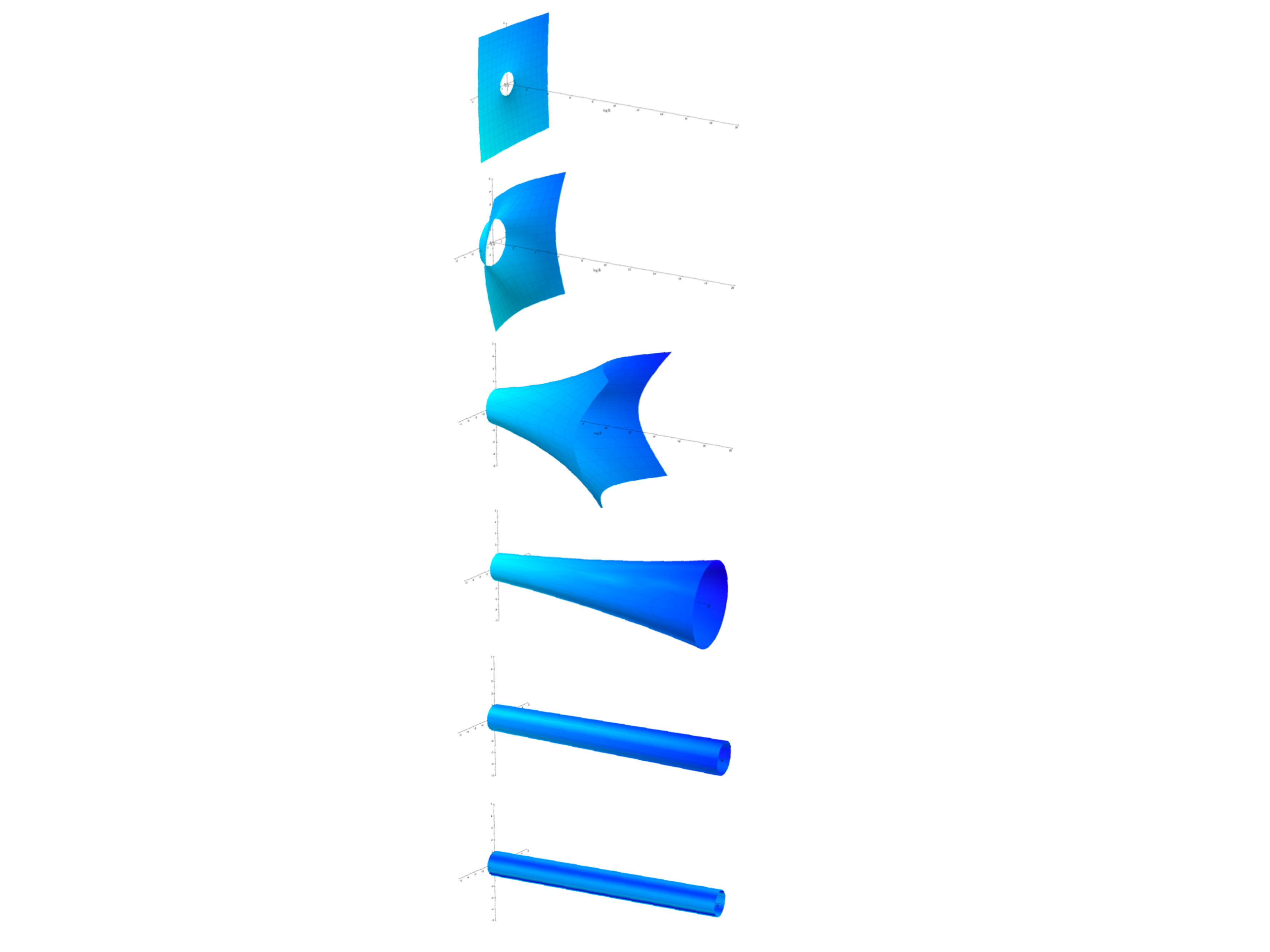}
\caption{The coordinates $Z$ and $T$ are supressed, while $R$ and $\protect\psi $ are embedded in a three-dimensional Euclidean space. The sequence of figures (from top to bottom) represents the evolution with increasing $\protect\mu _{K}$ of the dependence of the Euclidean radius (defined as $C\left( R;\protect\mu _{K}\right) /2\protect\pi $, the circumference of the circles over $2\protect\pi $, represented in the vertical plane) on the coordinate radius $R$ (represented on logarithmic scale on the horizontal axis). At small $\protect\mu _{K}$ (top figure) the metric is almost flat. With increasing $\protect\mu _{K}$ (lower figures) gravity bends spacetime into a funnel (with the tip at $R=0$), which eventually degenerates into a cylinder, with unit radius for $\protect\mu _{K}\rightarrow \infty $ (bottom figure). While for any $\protect\mu _{K}\neq 0$ the tip $R=0$ of the funnel represents a singularity, the larger hole appearing at the base of the funnel at small $R$ is but a numerical artefact arising from the lower theshold in the chosen range of $R$ to be represented.}
\label{Fig_movie}
\end{figure}

To illustrate the effect of increasing $\mu _{K}$, we calculate the
circumference of the circles with coordinate radius $R$ from Eq. (\ref%
{LCfinal}) as 
\begin{equation}
C\left( R;\mu _{K}\right) =2\pi \left[ \left( 1+2\mu _{K}\right) R\right] ^{%
\frac{1+\sqrt{1+8\mu _{K}}}{2\left( 1+2\mu _{K}\right) }}~.
\end{equation}%
Then we plot the circumference as function of both $R$ and $\mu _{K}$ on
Fig. \ref{circumference3D}. While for $\mu _{K}=0$ the circumference takes
the flat value $C\left( R;0\right) =2\pi R$, at $\mu _{K}\rightarrow \infty $
it becomes $\lim_{\mu _{K}\rightarrow \infty }C\left( R;\mu _{K}\right)
=2\pi \lim_{\mu _{K}\rightarrow \infty }\mu _{K}^{1/\sqrt{2\mu _{K}}}=2\pi $%
. We also represent the process of how the Euclidean radius of the circles
(defined as $C\left( R;\mu _{K}\right) /2\pi $) changes with the coordinate $%
R$ for increasing values of $\mu _{K}$ on both an animation (with increasing 
$\mu _{K}$ as time variable), given as supplementary material, and on a
sequence of figures (Fig. \ref{Fig_movie}.)

\begin{figure*}
\includegraphics[scale=0.5]{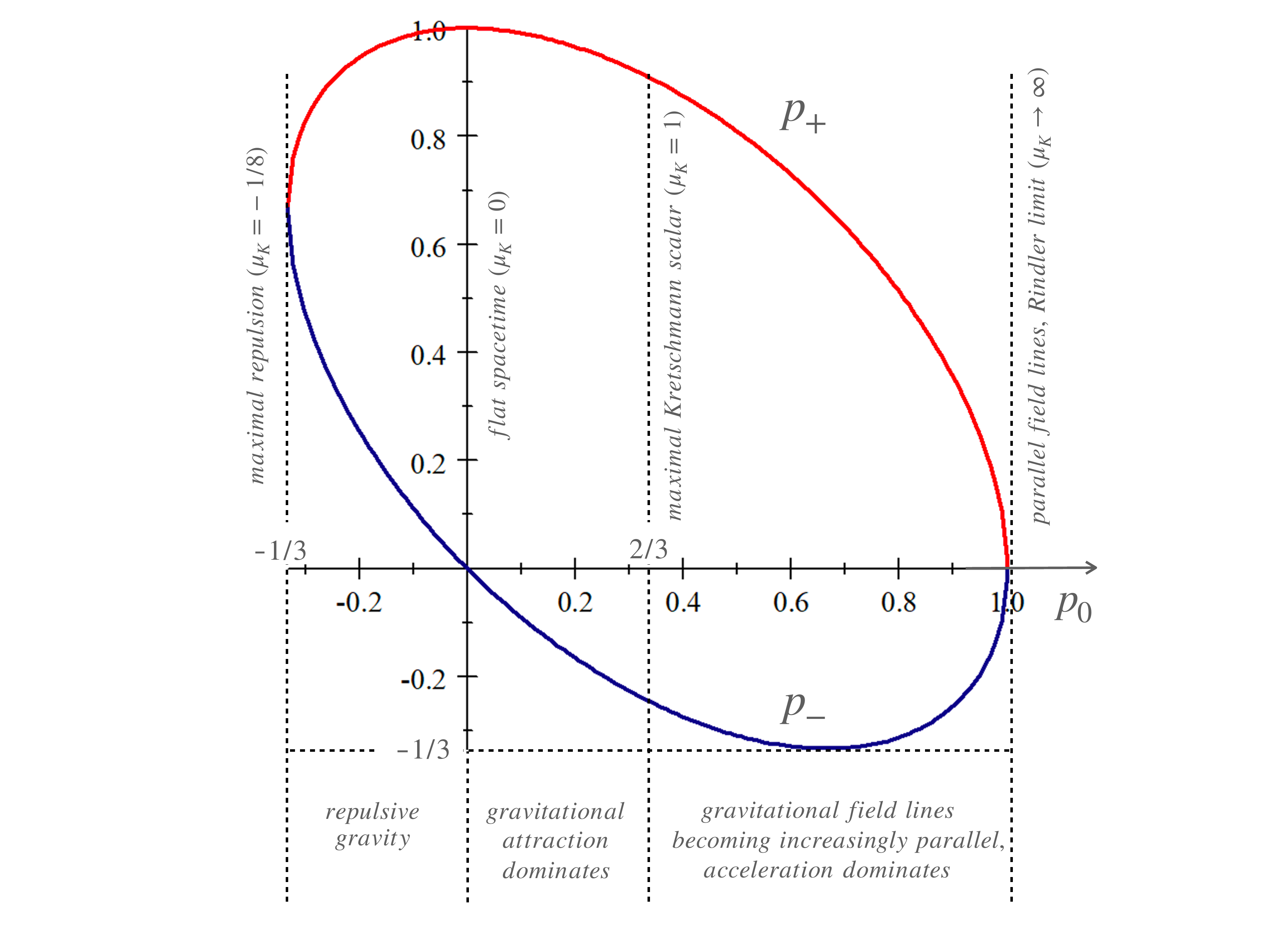}
\caption{The various regimes of the Levi-Civita metric in terms of the Kasner parameters $p_{+}$ (red curve), $p_{-}$ (blue curve) and $p_{0}$ (horizontal axis). The Komar mass density increases from left to right, spanning to the regime of negative gravity $\protect\mu _{K}\in \lbrack -1/8,0)$, no gravity $\protect\mu _{K}=0$, gravitational attraction dominated regime $\protect\mu _{K}\in (0,1)\,$, maximal Kretschmann scalar (the metric of Ref. \protect\cite{Ahmed}) $\protect\mu _{K}=1$, increasingly parallel field lines transforming gravity into an acceleration field $\protect\mu _{K}\in (1,\infty )$, and perfectly parallel field lines, the Rindler limit $\protect\mu _{K}\rightarrow \infty $.}
\label{Fig_regimes}
\end{figure*}

We summarize the findings of the paper by presenting the various regimes of
the Levi-Civita spacetime on Fig. \ref{Fig_regimes}. On the horizontal axis
the Kasner parameter $p_{0}$ increases from $-1/3$ to $1$. The red and blue
curves represent $p_{+}$ and $p_{-}$, respectively. The Komar mass density
also increases from left to right, monotonically with $p_{0}$. From left to
right the figure shows the following regimes:

\begin{itemize}
\item[i)] the limit of maximal repulsion, for $\mu _{K}=-1/8$, thus $%
p_{0}=-1/3$

\item[ii)] the repulsive gravity regime, for $\mu _{K}\in \lbrack -1/8,0)$,
thus $p_{0}\in \lbrack -1/3,0)$

\item[iii)] the flat limit, for $\mu _{K}=0$, thus $p_{0}=0$

\item[iv)] the regime, where gravitational attraction dominates, for $\mu
_{K}\in (0,1)$, thus $p_{0}\in (0,2/3)$

\item[v)] the maximal value of the Kretschmann scalar (the metric of Ref. 
\cite{Ahmed}), for $\mu _{K}=1$, thus $p_{0}=2/3$

\item[vi)] the regime, where Newtonian gravity drags the field lines
increasingly parallel, for $\mu _{K}\in (1,\infty )$, thus $p_{0}\in (2/3,1)$

\item[vii)] the Rindler limit, where the perfectly parallel field lines
transform gravity into a pure acceleration field through the equivalence
principle, for $\mu _{K}\rightarrow \infty $, thus $p_{0}=1$.
\end{itemize}

Hence, from the combined analysis of the gravitational acceleration (\ref{accG}), which increases monotonically with $\mu _{K}$ but asymptotes to a constant; and of the tidal force (which falls off completely in the asymptotic regime $\mu _{K}\rightarrow \infty $), as shown by the geodesic deviation acceleration (\ref{accDev}) we conclude, that adding to the Komar mass density strengthens the gravitational acceleration, as expected, however drives the field lines increasingly parallel. The first effect dominates at small $\mu _{K}$, while the second at large $\mu _{K}$. The Riemannian curvature decays with increasing $\mu _{K}$, the gravitational field becoming fully equivalent to a Rindler frame of accelerating observers in flat spacetime. Hence, in a Newtonian sense the field lines become parallel, while in an Einsteinian sense gravity vanishes.

\section{Acknowledgements}

This work was supported by the Hungarian National Research Development and
Innovation Office (NKFIH) in the form of Grant No. 123996 and has been
carried out in the framework of COST action CA18108 (QG-MM) supported by
COST (European Cooperation in Science and Technology).
Computations of curvature components were performed with the computer algebra system
Cadabra2 \cite{Cad1,Cad2}.

\ \appendix

\section{Derivation of the Einstein--Rosen form of the cylindrically
symmetric metric\label{stacan}}

Thorne \cite{Thorne} has given the line element for a generic cylindrically
symmetric spacetime with vorticity-free Killing vectors and orthogonally
transitive group action (dubbed as whole-cylinder symmetry) in a standard
form 
\begin{equation}
\mathrm{d}s^{2}=e^{2\left( \tilde{K}-U\right) }\left( -\mathrm{d}\tilde{t}%
^{2}+\mathrm{d}\tilde{r}^{2}\right) +e^{-2U}W^{2}\mathrm{d}\varphi
^{2}+e^{2U}\mathrm{d}z^{2}~,  \label{standard}
\end{equation}%
with $U$, $\tilde{K}$ and $W$ functions of $\left( \tilde{t},\tilde{r}%
\right) $ only. For certain particular sources this can be reduced to the
simpler Einstein--Rosen (canonical) form (\ref{canonical}). For the
completeness of presentation we discuss in this Appendix explicitly the
reduction and coordinate transformation leading to the canonical form of the
cylindrically symmetric line element.

As Thorne has emphasized, when the energy density equals the radial
pressure, thus $T_{\tilde{t}}^{\tilde{t}}+T_{\tilde{r}}^{\tilde{r}}=0\,$\ (a
condition holding both for vacuum and electromagnetic field), the Einstein
equation $R_{\tilde{t}}^{\tilde{t}}+R_{\tilde{r}}^{\tilde{r}}-R=0$ implies
that the function $W$ with differential 
\begin{equation}
\mathrm{d}W=\frac{\partial W}{\partial \tilde{t}}\mathrm{d}\tilde{t}+\frac{%
\partial W}{\partial \tilde{r}}\mathrm{d}\tilde{r}  \label{Wdef}
\end{equation}%
obeys the wave equation $W_{,\tilde{t}\tilde{t}}=W_{,\tilde{r}\tilde{r}}$.
Hence $W$ is a harmonic function, advantageous to use as a new coordinate.

We introduce another new coordinate $t$ through 
\begin{equation}
\frac{\partial t}{\partial \tilde{t}}=\frac{\partial W}{\partial \tilde{r}}%
~,\quad \frac{\partial t}{\partial \tilde{r}}=\frac{\partial W}{\partial 
\tilde{t}}~.  \label{tdef}
\end{equation}%
The new coordinate is well-defined, since the harmonicity of $W$ is but the
integrability condition for $t$. Moreover, $t$ is also harmonic through Eq. (%
\ref{tdef}). From Eqs. (\ref{Wdef}) and (\ref{tdef}) it is immediate to show
that%
\begin{equation}
-\mathrm{d}t^{2}+\mathrm{d}r^{2}=e^{2\alpha }\left( -\mathrm{d}\tilde{t}^{2}+%
\mathrm{d}\tilde{r}^{2}\right) ~,
\end{equation}%
with $\alpha $ a function of $\left( \tilde{t},\tilde{r}\right) $ given as%
\begin{equation}
e^{2\alpha }=-\left( \frac{\partial W}{\partial \tilde{t}}\right)
^{2}+\left( \frac{\partial W}{\partial \tilde{r}}\right) ^{2}~.
\label{eadal}
\end{equation}%
Hence the set of harmonic coordinates $\left( t,r\equiv W\right) $,
similarly to the old coordinates $\left( \tilde{t},\tilde{r}\right) $, are
conformally flat.

In defining $\alpha $ we assumed the right hand side of Eq. (\ref{eadal})
positive, implying the 4-gradient of $W$ to be spacelike. It follows that $t$
is a temporal coordinate.

Eq. (\ref{eadal}) is trivially satisfied by parametrizing the 4-gradient of $%
W$ through 
\begin{eqnarray}
\frac{\partial W}{\partial \tilde{r}} &=&e^{\alpha }\cosh \beta ~,  \notag \\
\frac{\partial W}{\partial \tilde{t}} &=&e^{\alpha }\sinh \beta ~,
\end{eqnarray}%
with $\beta $ a function of $\left( \tilde{t},\tilde{r}\right) $. The
coordinate transformation then emerges as a sequence of a hyperbolic
rotation and a dilation on the coordinate differentials: 
\begin{equation}
\left( 
\begin{array}{c}
dt \\ 
dr%
\end{array}%
\right) =e^{\alpha }\left( 
\begin{array}{cc}
\cosh \beta & \sinh \beta \\ 
\sinh \beta & \cosh \beta%
\end{array}%
\right) \left( 
\begin{array}{c}
d\tilde{t} \\ 
d\tilde{r}%
\end{array}%
\right) ~.
\end{equation}%
With $K=\tilde{K}-\alpha $ and $U$ functions of $\left( t,r\right) $ the
line element in the new coordinates takes the canonical (Einstein-Rosen)
form (\ref{canonical}).

\section{Komar mass density of the Rindler metric\label{KomarR}}

In this appendix we calculate a suitably defined Komar mass surface density
in flat spacetime, expressed in a reference frame uniformly accelerated into
the $x$ direction with acceleration $x^{-1}$, hence in Rindler coordinates: 
\begin{equation}
\mathrm{d}s^{2}=-x^{2}\mathrm{d}t^{2}+\mathrm{d}x^{2}+\mathrm{d}y^{2}+%
\mathrm{d}z^{2}~,
\end{equation}%
where $t$ is now the Rindler time. This metric covers the right quadrant of
the Minkowski spacetime, corresponding to $x>0$ in the Rindler coordinates
and has a coordinate singularity at the hyperplane $x=0$, which represents
an infinitely accelerated observer.

We define $y=l\varphi $ and $z=k\psi $, with $l$ and $k$ length scales, $%
0\leq \varphi \leq 2\pi $ and $0\leq \psi \leq 2\pi $ angular coordinates
with period $2\pi $. This compactifies the spacetime in both the $y$ and $z$
direction with the original spacetime recovered for $l,k\rightarrow \infty $%
. In terms of the compactified coordinates the line element becomes 
\begin{equation}
\mathrm{d}s^{2}=-x^{2}\mathrm{d}t^{2}+\mathrm{d}x^{2}+l^{2}\mathrm{d}\varphi
^{2}+k^{2}\mathrm{d}\psi ^{2}~.
\end{equation}%
The timelike Killing vector $\xi =\partial _{t}$ has the length squared $\xi
\cdot \xi =-x^{2}$ forbidding a distinguished normalization for $\xi $. The
Komar superpotential reads 
\begin{equation}
\mathbf{U}_{\xi }=\ast \mathrm{d}\boldsymbol{\xi }=-2kl\mathrm{d}\varphi
\wedge \mathrm{d}\psi ~,
\end{equation}%
which integrated on $t=\mathrm{const}$, $x=\mathrm{const}$, $0\leq \varphi
\leq 2\pi $, and $0\leq \psi \leq 2\pi $ gives 
\begin{equation}
m_{K}=\pi kl~.
\end{equation}%
The coordinate area of the torus $0\leq \varphi ,\psi \leq 2\pi $ in the
original coordinates $y,z$ is $4\pi ^{2}kl$, we thus define the surface
Komar mass density as 
\begin{equation}
\left( \sigma _{K}\right) _{_{\mathrm{Rindler}}}=\frac{m_{K}}{4\pi ^{2}kl}=%
\frac{1}{4\pi }.  \label{sigmaK}
\end{equation}%
We may now take $k,l\rightarrow \infty $ to recover the original Rindler
spacetime, a procedure which does not affect $\left( \sigma _{K}\right) _{_{%
\mathrm{Rindler}}}$.

With respect to the timelike coordinate vector of the standard
pseudo-Cartesian coordinates, Minkowski spacetime has zero Komar mass.
Indeed, the timelike Killing vector is naturally normalized everywhere.

When no obvious normalization of the Killing vector (ensured for example by
a proper asymptotic behaviour) is available, Komar integrals can lead to
finite, conserved charges that capture some aspects of the reference frame
(in this case, acceleration), but such charges do not necessarily
characterize invariant geometric aspects of the gravitational field.

\section{A particular case: the maximal Kretschmann parameter\label{AhmedNO}}

In this appendix we show that the cylindrically symmetric vacuum spacetime
discussed in Ref. \cite{Ahmed} is but a particular case of the Levi-Civita
metric. We also refute some of their claims about the spacetime.

We start by introducing new coordinates as%
\begin{eqnarray}
T &=&\left( \frac{3}{2}\right) ^{2/3}t_{\ast }~,\quad R=\frac{2}{3}\sinh
^{3/2}r_{\ast }~,  \notag \\
\chi &=&\left( \frac{3}{2}\right) ^{2/3}\varphi _{\ast }~,\quad \zeta
=\left( \frac{3}{2}\right) ^{-1/3}z_{\ast }~,
\end{eqnarray}%
(with $\chi \in \left[ 0,2\pi \left( 3/2\right) ^{2/3}\right] $), rendering
the line element (\ref{Farook}) into (\ref{LCKasner}), with the particular
coefficients $p_{0}=p_{+}=2/3$ and $p_{-}=-1/3$, leading to $\mu _{K}=1$.
Hence the metrics (\ref{LCER}) and (\ref{Farook}) are locally equivalent for
this particular parameter value, nevertheless there is a disagreement in the
angular deficits in $\chi $ in their Kasner form.

At $\mu _{K}=1$, the particular case of the Levi-Civita metric discussed in
Ref. \cite{Ahmed}, the Kretschmann scalar exhibits its maximum:%
\begin{equation}
\mathcal{K}=\left( \frac{4}{3}\right) ^{3}R^{-4}=\frac{12}{\sinh ^{6}r_{\ast
}}~
\end{equation}%
and the curvature invariants (\ref{scalars}) are%
\begin{eqnarray}
J_{1} &=&\mathcal{K~},\quad J_{2}=-\frac{\mathcal{K}^{3/2}}{2\sqrt{3}}=-%
\frac{12}{\sinh ^{9}r_{\ast }}~, \\
J_{3} &=&-\frac{\mathcal{K}^{2}}{4}=-\frac{36}{\sinh ^{12}r_{\ast }}~,\quad
J_{4}=0~.
\end{eqnarray}%
With this, we correct the values of $J_{2}$ and $J_{4}$ given in Ref. \cite%
{Ahmed}.

The C-energy (\ref{EC}) in the particular case $\mu _{K}=1$ reads%
\begin{equation}
E_{C}^{\mu _{K}=1}=\ln r~=\frac{\ln 2}{3}+\frac{1}{2}\ln \sinh r_{\ast },
\end{equation}%
also different from the one given in Ref. \cite{Ahmed}, which however seems
to be calculated from $E_{C}^{\mathrm{alt}}$.

Ref. \cite{Ahmed} additionally claimed that the singularity on the axis is
soft. We also refute this statement in section \ref{Sing} and Appendix \ref%
{StrongSing}.

\section{Strong singularity conditions\label{StrongSing}}

In this Appendix we give the details of the calculations of the Tipler and Kr%
ólak integrals necessary to classify the singularity on the axis.

In the special case when the Weyl tensor is not identically zero, does not
show oscillatory behaviour along the geodesic and the geodesic is null, a
unified necessary and sufficient condition for a strong singularity both in
the sense of Tipler (\ref{LTip}) and of Królak (\ref{NKro}) can be given, in
terms of the components $C_{\ bcd}^{a}$ of the Weyl tensor in a parallel
propagated pseudoorthonormal frame $\left\{ e_{\mathbf{a}}\right\} $, with
dual $\left\{ \theta ^{\mathbf{a}}\right\} $. Such a frame, with $e_{\mathbf{%
0}}$ the tangent of the (affinely parametrized) null geodesic (written in
terms of Kasner-like coordinates) was presented as Eq. (\ref{e03}) and can
be conveniently extended to a neighborhood of the geodesic.

The curvature forms $\Omega _{\ \mathbf{b}}^{\mathbf{a}}=\mathrm{d}\omega _{~%
\mathbf{b}}^{\mathbf{a}}+\omega _{~\mathbf{c}}^{\mathbf{a}}\wedge \omega _{~%
\mathbf{b}}^{\mathbf{c}}$ (with $\omega _{~\mathbf{b}}^{\mathbf{a}}$ the
connection 1-forms) are 
\begin{align}
\Omega _{\ \mathbf{0}}^{\mathbf{0}}& =\frac{p_{0}\left( p_{0}-1\right) }{%
R^{2}}\theta ^{\mathbf{0}}\wedge \theta ^{\mathbf{1}}~,  \notag \\
\Omega _{\ \mathbf{2}}^{\mathbf{0}}& =\frac{p_{+}p_{-}}{2R^{2}}\theta ^{%
\mathbf{0}}\wedge \theta ^{\mathbf{2}}-\frac{\left( 2p_{0}+p_{-}\right) p_{+}%
}{4R^{2\left( 1-p_{0}\right) }}\theta ^{\mathbf{1}}\wedge \theta ^{\mathbf{2}%
}~,  \notag \\
\Omega _{\ \mathbf{3}}^{\mathbf{0}}& =\frac{p_{+}p_{-}}{2R^{2}}\theta ^{%
\mathbf{0}}\wedge \theta ^{\mathbf{3}}-\frac{\left( 2p_{0}+p_{+}\right) p_{-}%
}{4R^{2\left( 1-p_{0}\right) }}\theta ^{\mathbf{1}}\wedge \theta ^{\mathbf{3}%
}~,  \notag \\
\Omega _{\ \mathbf{2}}^{\mathbf{1}}& =\frac{p_{+}p_{-}}{2R^{2}}\theta ^{%
\mathbf{1}}\wedge \theta ^{\mathbf{2}}-\frac{\left( 2p_{0}+p_{-}\right) p_{+}%
}{R^{2\left( 1+p_{0}\right) }}\theta ^{\mathbf{0}}\wedge \theta ^{\mathbf{2}%
}~,  \notag \\
\Omega _{\ \mathbf{3}}^{\mathbf{1}}& =\frac{p_{+}p_{-}}{2R^{2}}\theta ^{%
\mathbf{1}}\wedge \theta ^{\mathbf{3}}-\frac{\left( 2p_{0}+p_{+}\right) p_{-}%
}{R^{2\left( 1+p_{0}\right) }}\theta ^{\mathbf{0}}\wedge \theta ^{\mathbf{3}%
}~,  \notag \\
\Omega _{\ \mathbf{3}}^{\mathbf{2}}& =-\frac{p_{+}p_{-}}{R^{2}}\theta ^{%
\mathbf{2}}\wedge \theta ^{\mathbf{3}}~,
\end{align}%
with the frame indices raised and lowered by the flat metric 
\begin{equation}
\left( k_{ab}\right) =\left( k^{ab}\right) =%
\begin{pmatrix}
0 & -1 & 0 & 0 \\ 
-1 & 0 & 0 & 0 \\ 
0 & 0 & 1 & 0 \\ 
0 & 0 & 0 & 1%
\end{pmatrix}%
~,
\end{equation}%
which for an arbitrary vector $V^{a}$ results in the rules 
\begin{equation}
V_{0}=-V^{1}~,\quad V_{1}=-V^{0}~,\quad V_{2}=V^{2}~,\quad V_{3}=V^{3}~.
\end{equation}

The Levi-Civita spacetime being Ricci-flat, the curvature forms $\Omega _{\ 
\mathbf{b}}^{\mathbf{a}}=\frac{1}{2}R_{\ bcd}^{a}\theta ^{\mathbf{c}}\wedge
\theta ^{\mathbf{d}}$ represent Weyl tensor components, nonvanishing along
the geodesic (except for special parameter values for which the spacetime is
flat, but then there is no singularity either), also they do not oscillate.
Hence the criteria for the diverging of the Tipler and Królak integrals
representing unified necessary and sufficient conditions for a strong
singularity are met.

Except the flat case, thus either of the cases $\left(
p_{0},p_{+},p_{-}\right) =\left( 0,1,0\right) $ or $\left( 1,0,0\right) $,
all components of the Weyl tensor blow up with $R\rightarrow 0$. In what
follows, we discuss, whether this singularity is strong or soft.

The explicit calculation gives the nonvanishing Królak integrals 
\begin{equation}
N_{\ 1}^{0}=D-CR^{2p_{0}}-\frac{C^{2}}{p_{0}+1}R^{p_{0}+1}-p_{0}^{2}F_{1}%
\left( R;p_{0}\right) ~,
\end{equation}%
where $C$ and $D$ are constants of integration, and 
\begin{equation}
F_{1}\left( R;p_{0}\right) =%
\begin{cases}
\frac{R^{3p_{0}-1}}{3p_{0}-1}~, & p_{0}\neq 1/3 \\ 
\ln R~, & p_{0}=1/3%
\end{cases}%
~,
\end{equation}%
together with%
\begin{align}
N_{\ 2}^{2}& =D+\frac{p_{+}^{2}\left( 2p_{0}+p_{-}\right) ^{2}}{\left(
p_{0}+1\right) ^{3}}R^{-\left( p_{0}+1\right) }-\frac{C^{2}}{p_{0}+1}%
R^{p_{0}+1}  \notag \\
& -2C\frac{p_{+}\left( 2p_{0}+p_{-}\right) }{p_{0}+1}\ln R~,  \notag \\
N_{\ 3}^{3}& =D+\frac{p_{-}^{2}\left( 2p_{0}+p_{+}\right) ^{2}}{\left(
p_{0}+1\right) ^{3}}R^{-\left( p_{0}+1\right) }-\frac{C^{2}}{p_{0}+1}%
R^{p_{0}+1}  \notag \\
& -2C\frac{p_{-}\left( 2p_{0}+p_{+}\right) }{p_{0}+1}\ln R~.
\end{align}%
At $R\rightarrow 0$, the integrals $N_{\ 2}^{2}$ and $N_{\ 3}^{3}$ contain
both power-law and logarithmic divergences, while $N_{\ 1}^{0}$ can blow up
for any $p_{0}\leq 1/3$ parameter values.

Likewise, the Tipler integrals $L_{\ b}^{a}$ are%
\begin{eqnarray}
L_{\ 1}^{0} &=&E-\frac{D}{p_{0}+1}R^{p_{0}+1}+\frac{C}{3p_{0}+1}R^{3p_{0}+1}
\notag \\
&&+\frac{C^{2}R^{2p_{0}+2}}{2\left( p_{0}+1\right) ^{2}}+\frac{p_{0}}{4}%
R^{4/3}F_{2}\left( R;p_{0}\right) ~,
\end{eqnarray}%
where $C,D$ and $E$ are constants of integration, and 
\begin{equation}
F_{2}\left( R;p_{0}\right) =%
\begin{cases}
\frac{R^{4\left( 3p_{0}-1\right) /3}}{3p_{0}-1}~, & p_{0}\neq 1/3 \\ 
\ln R~, & p_{0}=1/3%
\end{cases}%
~,
\end{equation}%
together with 
\begin{eqnarray}
L_{\ 2}^{2} &=&\int_{R}^{R_{0}}\mathrm{d}R\,\left[ DR^{p_{0}}+\frac{\left(
2p_{0}+p_{-}\right) ^{2}p_{+}^{2}}{\left( p_{0}+1\right) ^{3}R}-\frac{%
C^{2}R^{2p_{0}+1}}{p_{0}+1}\right.  \notag \\
&&\left. -2C\frac{\left( 2p_{0}+p_{-}\right) p_{+}}{p_{0}+1}R^{p_{0}}\ln R%
\right] ~,
\end{eqnarray}%
(with $R_{0}$ characterizing the initial point of the geodesic) and $L_{\
3}^{3}$ obtained from $L_{\ 2}^{2}$ through the exchange $%
p_{+}\leftrightarrow p_{-}$. We can see even without calculating the
integrals a logarithmic divergence emerging in both $L_{\ 2}^{2}$ and $L_{\
3}^{3}$ at $R\rightarrow 0$, while $L_{\ 1}^{0}$ blows up only for $%
p_{0}\leq 1/3$.
We conclude that for radial null geodesics the strong singularity conditions
are satisfied both in the sense of Królak and of Tipler.

\end{document}